\documentclass[twocolumn, usenatbib]{mnras}
\usepackage{savesym}
\usepackage{graphicx}
\usepackage{longtable}
\usepackage{changepage}

\expandafter\let\csname equation*\endcsname\relax
  \expandafter\let\csname endequation*\endcsname\relax 
\usepackage{subfig}
\usepackage{amsmath}
\usepackage{amssymb}
\usepackage{verbatim}
\usepackage[yyyymmdd,hhmmss]{datetime}
\usepackage{array}
\usepackage{times}
\usepackage{xcolor}

\DeclareRobustCommand{\DE}[3]{#2}
\let\DEthebibliography\thebibliography
\def\thebibliography{\DeclareRobustCommand{\DE}[3]{##3}\DEthebibliography}

\newcommand{\beq}{\begin{equation}}
\newcommand{\eeq}{\end{equation}}

\newcommand\sk{{{\tt superkerr}\ }}

\title[Super-extremal Kerr discs]{ Testing theories of accretion and gravity with super-extremal Kerr discs }
\author [Andrew Mummery, Steven Balbus, Adam Ingram]{Andrew Mummery$^1$\thanks{E-mail:
andrew.mummery@physics.ox.ac.uk}, Steven Balbus$^1$, Adam Ingram$^2$
\\
$^1$Oxford Theoretical Physics, Beecroft Building,  Clarendon Laboratory, Parks Road, Oxford, OX1 3PU, United Kingdom\\
$^2$School of Mathematics, Statistics and Physics, Newcastle University, Herschel Building, Newcastle upon Tyne, NE1 7RU, UK }

\date{}

\begin{document}

\pagerange{\pageref{firstpage}--\pageref{lastpage}} \pubyear{2023}

\maketitle

\label{firstpage}

\begin{abstract} 
 Fitting the thermal continuum emission of accreting black holes observed across X-ray bands represents one of the principle means of constraining the properties (mass and spin) of astrophysical black holes. Recent “continuum fitting” studies of Galactic X-ray binaries in the soft state have found best fitting dimensionless spin values which run into the prior bounds placed on traditional models ($a_\star = 0.9999$). It is of critical importance that these results are robust, and not a result solely of the presence of these prior bounds and deficiencies in conventional models of accretion. Motivated by these results we derive and present {\tt superkerr}, an {\tt XSPEC} model comprising of a thin accretion disc solution valid in the Kerr geometry for arbitrary spin parameter $a_\star$, extending previous models valid only for black holes ($|a_\star| < 1$).  This extension into “superextremal” spacetimes with $|a_\star| > 1$ includes solutions which describe discs evolving around naked singularities, not black holes. While being valid solutions of Einstein's field equations these naked singularities are not expected to be present in nature.  We discuss how the “measurement” of a Kerr spin parameter $1 < a_\star < 5/3$ would present compelling evidence for the requirement of a rethink in either standard accretion theory, or our theories of gravity. 
\end{abstract}

\noindent

\section{Introduction} 
The study of astrophysical black holes which accrete material from a binary companion represents {one of} the original observational probes of the strong field regime of gravity.  The accretion flows surrounding these objects evolve deep into the gravitational potential well of the central black hole, with their emitted X-ray photons carrying characteristic observational signatures of a highly relativistic spacetime.  Even with the advent of multi messenger probes of gravity \citep[e.g.,][]{Abbott16}, X-ray spectroscopy remains one of the principal techniques through which the physical properties (the mass and rate of rotation) of astrophysical black holes are probed. 

One such observational technique is named “continuum fitting”, and involves fitting the broad band spectral energy distribution (SED) observed from accreting systems across an X-ray bandpass.  As the details of the disc temperature profile, as well as various gravitational optics effects, are sensitive to both the mass and spin of the central black hole, these broad-band SEDs in principle contain sufficient information to constrain the black holes characteristics.  This technique has been employed widely throughout the literature \citep[e.g.,][]{Shafee06, Steiner09, McClintock14}. 

Of the two parameters which describes a Kerr spacetime, the spin parameter $a_\star$ is of considerable astrophysical interest. Representing the {sum of the natal spin and the} integrated angular momentum flux onto the black hole over its lifetime, the spin encodes information regarding the formation and evolution of astrophysical black holes.  To substantially change the spin of a black hole via the accretion of material (which naturally carries angular momentum across the event horizon of the black hole) an order unity fraction of the original black hole mass must be accreted \citep{Bardeen1970,Thorne1974}.  Low mass X-ray binaries (those with binary companion mass $\ll$ the black hole mass $M$) have insufficient mass to substantially modify the black hole's spin, while high mass X-ray binaries (those with binary companion masses $\gg$ than the black hole mass) last for short periods of time (owing to the short lifespans of high mass stars). These short lifespans mean that there is insufficient time for an Eddington limited accretion flow to accrete sufficient material to significantly change the black holes spin \citep[the][timescale $t_S \sim 5\times10^7$ years sets the relevant timescale for mass change via Eddington limited accretion]{Salpeter64}. Therefore Galactic X-ray binaries should have spin parameters which probe their “birth” spins, and likely therefore provide interesting insights into the deaths of massive stars \citep{King16}.

A recent analysis, using conventional approaches, by \citet{Zhao21} is therefore of real interest. \citet{Zhao21} analysed the continuum emission of Cygnus-X1, finding an extremely rapid rotation rate $a_\star > 0.9985$ at 3$\sigma$. In fact, the best fitting spin in the \citet{Zhao21} analysis ran into the bounds placed on conventional models $a_\star = 0.9999$. Even the 3$\sigma$ lower limit found by \citet{Zhao21} is difficult to reconcile with conventional theory, as it exceeds the \citet{Thorne1974} limit imposed by photon capture of $a_\star = 0.998$. The \citet{Thorne1974} limit is a natural consequence of the very models used to derive the \citet{Zhao21} spin constraints, and so this result is a concern for conventional theoretical models. 

Prior limited spins also occur for supermassive black hole systems modelled with the continuum fitting approach \citep{Wen20}. Spin constraints from continuum fitting analyses are of course model dependent, in that they require physical models for the properties of the underlying accretion flow to be specified. It is possible therefore that the spin constraints derived for Cyg-X1 may be in some part a result of modelling deficiencies in conventional accretion theory. {Naturally, analytical disc models are simplified in various ways, and comparisons between full numerical GRMHD simulations and analytical models of accretion flows do show systematic parameter biases \citep[e.g.,][for a recent example]{Lancova23}. Thin disc models are also known to suffer from formal viscous \citet{LightmanEardley74}, and thermal \citet{SS76} instabilities, although whether these instabilities manifest themselves in nature is a topic of some debate. Thin disc models do however provide formally good fits to observations of many accreting sources and this fact, coupled with their flexibility, means that thin disc models remain the dominant theoretical tool for analysing the thermal emission observed from accreting sources. Extensions of thin disc models to new regimes is therefore of real interest. }

One notable omission in classical {thin disc} models is the ignored  emission sourced at and within the ISCO (innermost stable circular orbit) of a black hole's spacetime.  Classical models impose the so-called `vanishing ISCO stress' boundary condition, which forces the temperature of the accretion flow to be zero at the ISCO, with no emission assumed to be produced between the ISCO and event horizon. However, numerical simulations \citep[e.g.,][]{Noble10, Zhu12, Schnittman16, Lancova19, Wielgus22} generically find non-zero temperatures at and within the ISCO, with non-zero associated thermal emission. This emission will modify the SED of an observed black hole disc, and may induce systematic errors in inferred black hole properties.  Recently, \citet{MummeryBalbus2023} extended classical accretion models through the ISCO, deriving analytical expressions for the disc thermodynamic properties valid at all radii.  These solutions will allow for the future testing of degeneracies between black hole properties and intra-ISCO emission. 

In this paper we take a different approach, and extend classical accretion models in a different sense. Motivated by the \citet{Zhao21} finding of prior-limited spins ($a_\star = 0.9999$), we extend classical models for Kerr metric discs into the super-extremal regime. The super-extremal regime is characterised by spin parameters $|a_\star|>1$.  This will facilitate the important test that (sub) extremal spins remain the best-fitting parameter values even when the entire parameter space (of Kerr metric spins) is explored by fitting procedures. 

The super-extremal limit (objects sometimes referred to as ``super-spinars'' in the literature) is accompanied by a qualitative change in behaviour of the metric, namely the event horizon of the Kerr metric {no longer exists} for all spin parameters $|a_\star| > 1$.  As such, the curvature singularity at the centre of the metric is ``naked’’, and is not shielded from the spacetime asymptotically far away.  We stress that Kerr metrics with $|a_\star| > 1$ remain perfectly valid mathematical solutions of Einstein's field equations (and so are in some sense physical), but that these metrics are not expected to be present in astrophysical sources.  This assumption (that all solutions of Einstein's equations which occur {under generic physical conditions} and which host singularities have those singularities hidden behind an event horizon) is known as the Weak Cosmic Censorship Conjecture \citep{Penrose1969}, and while unproven does have some circumstantial supporting evidence \citep{Wald97}. {Note that under certain contrived initial conditions \citep[e.g., the collapse of an infinite cylinder][or highly prolate gas spheroids  \citealt{Shapiro91}]{Chrusciel90} naked singularities are known to form. In this paper we are concerned more specifically with astrophysically realistic gravitational collapse and evolution (as we shall be modelling astrophysical X-ray sources) and therefore, for the following reasons, consider it unlikely that naked singularities will be the central objects in observable X-ray sources.    }

Firstly, it appears extremely difficult (or potentially impossible) to form a naked Kerr singularity in an astrophysically realistic setting \citep[][although not, as we have mentioned, under contrived initial conditions]{Giacomazzo11}. One also cannot, for example, fire a test particle into a black hole and over-spin it \citep{Wald74, Barausse10}, nor can a black hole be over-spun by accretion, which saturates at \citep[or in fact just below][]{Thorne1974} the extremal bound \citep{Bardeen1970}.  In addition, unlike black hole solutions which are known to be stable to matter perturbations \citep[][and references therein]{Klainerman21}, Kerr like naked singularities appear to be unstable to perturbations \citep{Dotti08, Pani10} and so may well rapidly transform back into black hole solutions if formed. Accretion onto a Kerr naked singularity is also known to spin-down the object until it reaches the extremal bound and becomes a black hole \citep{deFelice78}. 

The concept of using accreting systems to place bounds on the weak cosmic censorship conjecture has a long history (for a recent example see \cite{Vieira23}). The evolutionary effects of thin disc accretion on naked singularities \citep{deFelice78}, and some properties of the discs themselves \citep{Takashi10, Kovacs10} have been studied in the literature.  Some {\tt XSPEC} models which use modified gravity extensions of the Kerr metric are available, for both continuum fitting \citep[see][for the {\tt nkbb} model]{nkbbref} and iron line analysis \citep[see][for the {\tt relxill\_nk} model]{relxillnkref}.   These models make use of the Johannes-Psaltis metric \citep{Johannsen11}, which is not a vacuum solution of Einsteins equations (and therefore these approaches represent a modified gravity theory). 

As far as we are aware, no general discussion and {\tt XSPEC} implementation of super-extremal Kerr discs exists in the literature.  A full {\tt XSPEC} \citep{Arnaud96} implementation is crucial for testing theories of accretion and gravity, as most modern X-ray data analysis takes place using {\tt XSPEC} functionality.  

In this paper we present {\tt superkerr}\footnote{\url{https://github.com/andymummeryastro/superkerr}}, a full {\tt XSPEC} implementation of thin disc solutions valid in the Kerr geometry for arbitrary spin parameter $a_\star$, extending previous models valid only for black holes ($|a_\star| < 1$).  Our model includes all relevant gravitational optics via numerical ray tracing calculations, and allows smooth transitions from black hole ($|a_\star| < 1$) to naked singularity ($|a_\star| > 1$) spacetimes. 

We then discuss degeneracies between vanishing ISCO stress black hole and naked singularity disc systems for certain regions of $a_\star > 1$ parameter space \citep{Takashi10}, before focussing on a region of parameter space $1 < a_\star < 5/3$ where no such degeneracy exists. We highlight how black hole disc systems which have a non-zero temperature at the ISCO look like vanishing ISCO stress naked singularity disc systems in this key $1 < a_\star < 5/3$ part of parameter space. When restricted to black hole only solutions these finite ISCO temperature solutions return spins pegged at the imposed prior boundary $a_\star = 0.9999$, much like recent observations of Cyg-X1 \citep{Zhao21}.  Super-extremal disc solutions therefore offer a potentially powerful probe of disc physics. 

We also show that if super-extremal Kerr singularities do exist in nature, and have spins $1 < a_\star < 5/3$ then for the vast majority of this parameter space there exists no degeneracy with black hole systems, with or without a non-zero ISCO temperature. These models therefore allow us to prove our theories of gravity and search for such exotic objects.

In this paper we will present a complete analysis of thin disc accretion around naked singularities. Not every result in this paper is new, but as these objects are not familiar to much of the X-ray astronomy community we believe it is important that a self contained analysis is presented. The layout of the paper is as follows. In section 2 we present an overview of the basic properties of the super-extremal Kerr metric, focusing on those properties most relevant for accretion (e.g., the properties of the circular orbits of test particles). In section 3 we discuss how a naked singularity evolves in the presence of an accretion flow. In section 4 we derive the temperature profiles of thin discs evolving in the super-extremal Kerr metric. In section 5 we discuss photon orbits in the super-extremal Kerr metric, and the image plane properties of naked singularity discs. In section 6 we present our analysis of super-extremal Kerr disc X-ray spectra, before concluding in section 7.

\section{Super-extremal Kerr singularities   }
\subsection{The metric and its basic properties} 
Throughout this paper we will be examining discs evolving in the Kerr metric.    Our notation is as follows.   We use physical units in which we denote   the speed of light by  $c$, and the Newtonian gravitational constant by  $G$.    In coorindates $x^\mu$, the invariant line element ${\rm d}\tau$ is given by
\beq
{\rm d}\tau^2 = - g_{\mu\nu} {\rm d}x^\mu {\rm d}x^\nu,
\eeq
where $g_{\mu\nu}$ is the usual covariant metric tensor with spacetime indices $\mu, \nu$.   {The coordinates are standard $(t, r, \phi, z)$ Boyer-Lindquist in their near-equator form}:  $t$ is time as measured at infinity and the other symbols have their usual quasi-cylindrical interpretation.    

{In this paper, in common with previous analytical treatments of black hole discs, we shall assume that the fluid flow evolves in the Kerr midplane $\theta =\pi/2$. As discussed in the introduction, it is not necessarily the case that the fluid sourced from the companion star is aligned with the Kerr metric equatorial plane at large radii, as the direction of the compact object's spin axis may not have been significantly modified by historic accretion.  This is an assumption of convenience, as it substantially simplifies the algebraic disc equations}.   For metric mass parameter $M$, and angular momentum parameter $a$ (which has dimensions of length and is {\it unrestricted} in this work), the non-vanishing $g_{\mu\nu}$ metric components  are:
\begin{align}
g_{tt} &= -(1 - 2r_g/r)c^2 , \quad g_{t\phi} = g_{\phi t} = -2r_gac/r,  \nonumber \\
g_{\phi\phi} &= r^2+a^2 +2r_ga^2/r, \quad g_{rr} = r^2/\Delta,  \quad g_{zz} = 1, \nonumber \\ 
 \Delta &\equiv r^2 - 2r_g r +a^2, \quad \sqrt{-g} = r, \quad r_g \equiv GM/c^2. 
\end{align}
Note that our choice of coordinates has asymptotic (large radius) metric signature $(-1, +1, +1, +1)$. 

\subsubsection{(A lack of an) Event horizon }
The outer event horizon of the Kerr metric $r_+$ is defined as the larger of the two radii at which 
\beq
{1 \over g_{rr}(r_+)} = 0 \to r_+^2 - 2r_g r_+ + a^2 = 0. 
\eeq
The solution of this condition is trivial 
\beq
r_+ = r_g + \sqrt{r_g^2 - a^2} ,
\eeq
but highlights a crucial point of physics, namely that if 
\beq
|a| > r_g ,
\eeq
then the Kerr metric has no event horizon. This of course does not mean that the metric  ceases to be a solution of the governing equations of general relativity, merely that the singularity at $r=0$ is not ``hidden'' behind an event horizon.  As discussed in the introduction, this ``naked singularity" behaviour is hypothesised to {not form under generic initial conditions of gravitational collapse} \citep{Penrose1969}.

\subsection{Orbital motion and super-extremal ISCO's }
The key orbital component relevant for the study of thin accretion flows is the angular momentum of a fluid element undergoing circular motion, as this is an excellent approximation to the dynamical behaviour of the fluid in the stable disc regions. For the Kerr metric standard methods \citep[e.g.,][]{Hobson06} lead to the following circular angular momentum profile   
\beq\label{orbang}
U_\phi = (GMr)^{1/2}{(1+a^2/r^2-2ar_g^{1/2}/r^{3/2}) \over \sqrt{ 1 -3r_g/r +2ar_g^{1/2}/r^{3/2}}} .
\eeq
The (dimensionless) energy of a circular orbit is similarly given by
\beq\label{orben}
U_t/c^2 = - { (1-2r_g/r +ar_g^{1/2}/r^{3/2}) \over \sqrt{ 1 -3r_g/r +2ar_g^{1/2}/r^{3/2}} }.
\eeq
{In this paper we shall define the $\phi$ coordinate to increase in the direction of the orbital motion of the fluid.  This means that $U_\phi(r, a) \geq 0$ for all orbits, and retrograde motion is identified with negative Kerr spin parameters $a < 0$. }

\begin{figure}
\centering
\includegraphics[width=0.5\textwidth]{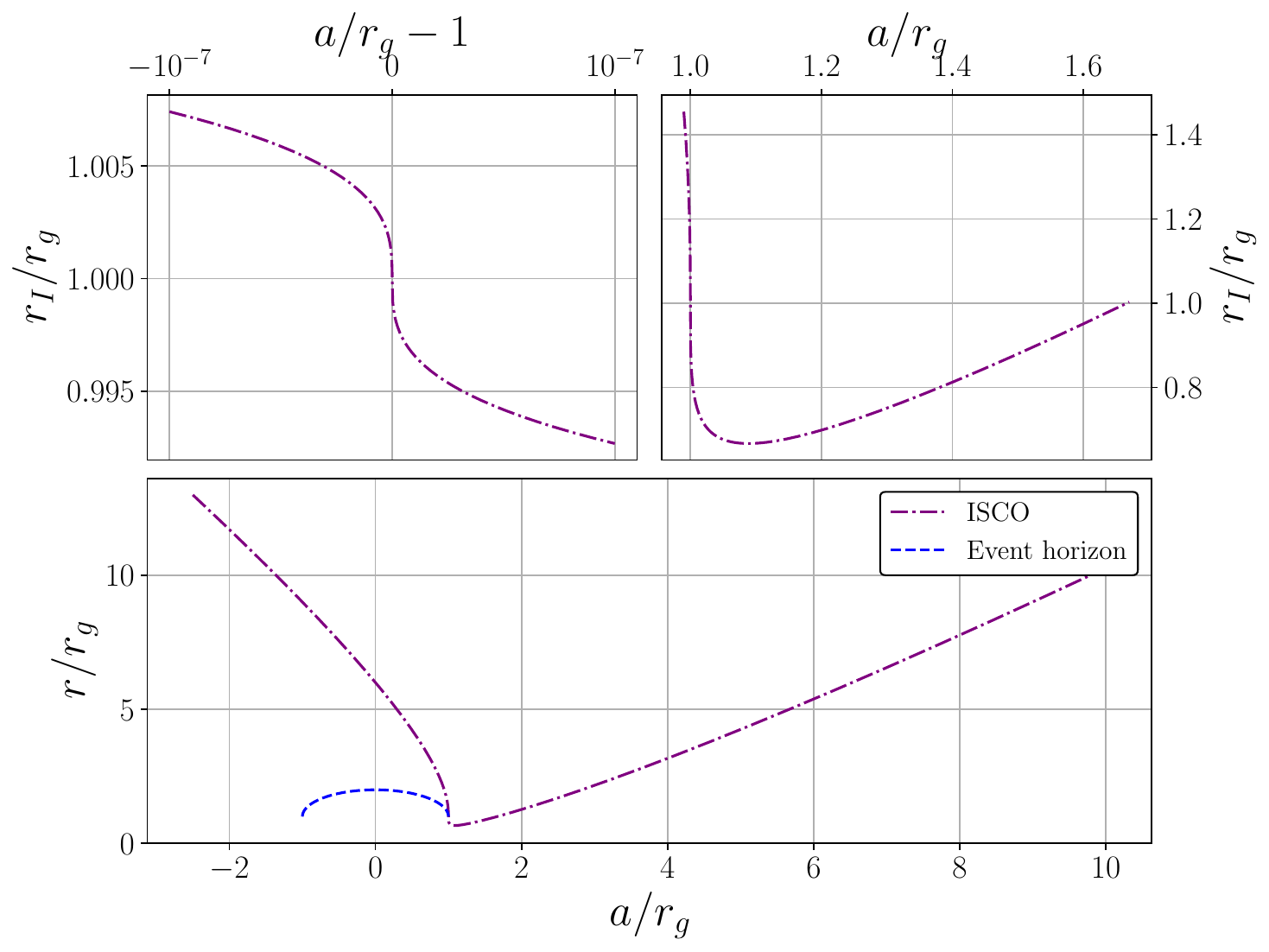}
\caption{Lower: The ISCO and event horizon locations as a function of Kerr metric spin parameter.  The ISCO is displayed by a purple dot-dashed curve, while the event horizon is shown by the blue dashed curve. The event horizon only exists for $|a_\star| <1$, while the ISCO exists for all spin parameters. The upper left panel shows the values of the ISCO for spin parameters close to $a_\star = 1$, while the upper right panel shows the turning point behaviour of the ISCO between $1 < a/r_g < 5/3$.   }
\label{fig1}
\end{figure}

These orbital solutions are valid in both the sub ($|a| < r_g$) and super ($|a| > r_g$) extremal regimes. 
A thin disc will persist in its classical form wherever these orbits are stable, and will transition onto rapid radial plunges where these orbits are unstable \citep[see][for an extension of classical accretion models into the plunging regime]{MummeryBalbus2023}. The stability of circular orbits  requires $\partial_r U_\phi > 0$, the relativistic analogue of the Rayleigh criterion.   Evaluating this gradient, we find 
\begin{equation}\label{grad_j}
{\partial U_\phi \over \partial r}  =
{ (GM)^{1/2}\over 2r^{1/2}{\cal D}}\left(1 +{ar_g^{1/2}\over r^{3/2}}\right)  \left( 1 - {6r_g\over r} -{3a^2\over r^2} +{8ar_g^{1/2}\over r^{3/2}}\right) ,
\end{equation}
where 
\beq
{\cal D} = \left( 1 -3r_g/r +2ar_g^{1/2}/r^{3/2}\right)^{3/2}.
\eeq 
Note that ${\cal D}$ is finite and positive for all radii $r > r_I$ of interest {(where we denote the ISCO location by $r_I$)}, so equation (\ref{grad_j}) passes through zero at $r=r_I$,  where the final factor vanishes, i.e.:
\begin{equation}\label{a}
r_I^2  -    6r_gr_I + 8a\sqrt{r_gr_I} - 3a^2  = 0.
\end{equation}
It is most illuminating to examine the spin as a function of the ISCO, as this involves solving a simple quadratic 
\beq
a = {4 \over 3} \sqrt{r_Ir_g} \pm \sqrt{{1\over 3}r_I^2 - {2\over 9} r_gr_I} .
\eeq
We see that {\it all spin parameters of the super-extremal Kerr metric have an ISCO}, with the ISCO's minimum radius being 
\beq
r_{I, {\rm min}} = {2 \over 3}r_g , 
\eeq
for the spin value 
\beq
a_{\rm min} = {2^{5/2} \over 3^{3/2}} r_g \simeq 1.089 r_g. 
\eeq
The ISCO location plotted as a function of Kerr metric spin parameter is displayed in Fig. \ref{fig1}. 

The location of the ISCO as a function of Kerr spin $r_I(a)$ can be written in closed form, and is given by \citep[cf.][]{Bardeen72} 
\beq
{r_I \over r_g}  = 3 + Z_2  - {\rm sgn}(a_\star) \sqrt{(3 - Z_1)(3 + Z_1 + 2 Z_2)},
\eeq
where 
\beq
a_\star \equiv a / r_g,
\eeq
is the dimensionless spin, and we have defined 
\beq
Z_1 = 
\begin{cases}
1 + (1-a_\star^2)^{1\over3}  \left((1+a_\star)^{1\over3} + (1-a_\star)^{1\over3}\right) , \quad  |a_\star| \leq 1 \\
\\
1 - (a_\star^2-1)^{1\over3} \left((1+|a_\star|)^{1\over3} - (|a_\star|-1)^{1\over3}\right) , \quad |a_\star| > 1,
\end{cases}
\eeq
and
\beq
Z_2 = \sqrt{3a_\star^2 + Z_1^2} .
\eeq
It will be important to note for future sections a set of interesting spin values. The first is $a_\star = 8\sqrt{6}/3$ for which $r_I = 6r_g$ just as in the case of a Schwarzschild black hole, the second $a_\star = 5/3$, for which $r_I = r_g$ just as in the case of an extremal (prograde) Kerr black hole, and finally $a_\star = 9$, for which $r_I = 9r_g$ just as in the case of an extremal (retrograde) Kerr black hole. Every Kerr black hole with an ISCO therefore has a corresponding super-extremal naked singularity with ISCO at the same physical location, with spin in the range $5/3 < a_\star < 9$. This spin range will be relevant for the study of the X-ray spectra emergent from these systems.  

\subsubsection{The ISCO angular momentum and energy }

The dimensionless energy of a circular orbit at the ISCO  radius is given by
\beq
\gamma_I(a) \equiv -U_t(r_I, a)/c^2 =  { (1-2r_g/r_I +ar_g^{1/2}/r_I^{3/2}) \over \sqrt{ 1 -3r_g/r_I +2ar_g^{1/2}/r_I^{3/2}} }.
\eeq
This quantity plays an important role in accretion theory, as the accretion efficiency (the amount of energy liberated from the fluid element over its full journey through the accretion flow, in units of its rest mass) is 
\beq
\eta(a) \equiv \gamma(r\to \infty) - \gamma_I = 1- \gamma_I . 
\eeq
Simplifying, we find
\begin{align}
\gamma_I &=  {1\over r_I} { (r_I^2 -2r_gr_I +ar_g^{1/2}r_I^{1/2}) \over \sqrt{ r_I^2 -3r_gr_I +2ar_g^{1/2}r_I^{1/2}} } ,\\
&= {1\over r_I} { (4r_gr_I -7ar_g^{1/2}r_I^{1/2}+3a^2) \over \sqrt{  3r_gr_I -6ar_g^{1/2}r_I^{1/2} + 3a^2} } ,\\
&= {4\over\sqrt{3}}\left(r_g\over r_I\right)^{1/2} \left( 1 -{3a\over 4\sqrt{r_gr_I}}\right) {\rm sgn}(\sqrt{r_I r_g}
- a) , \label{isco_energy}
\end{align}
where we have used the definition of the ISCO radius in going from the first to second lines, and have factorised the upper and lower quadratics in going to the final line. Note that the sign of the quantity $r_I^{1/2} r_g^{1/2} - a$ plays an important role in the dimensionless energy, and the energy of the last stable circular orbit is in fact discontinuous as $a_\star$ crosses the extremal bound $a_\star = 1$ into the super-extremal regime. A similar calculation for the ISCO angular momentum gives 
\beq
J_I = 2\sqrt{3}r_g c \left( 1 - {2a\over 3\sqrt{r_gr_I} }\right) {\rm sgn}(\sqrt{r_I r_g}
- a)  .\label{isco_ang_mom}
\eeq
\begin{figure}
\includegraphics[width=0.45\textwidth]{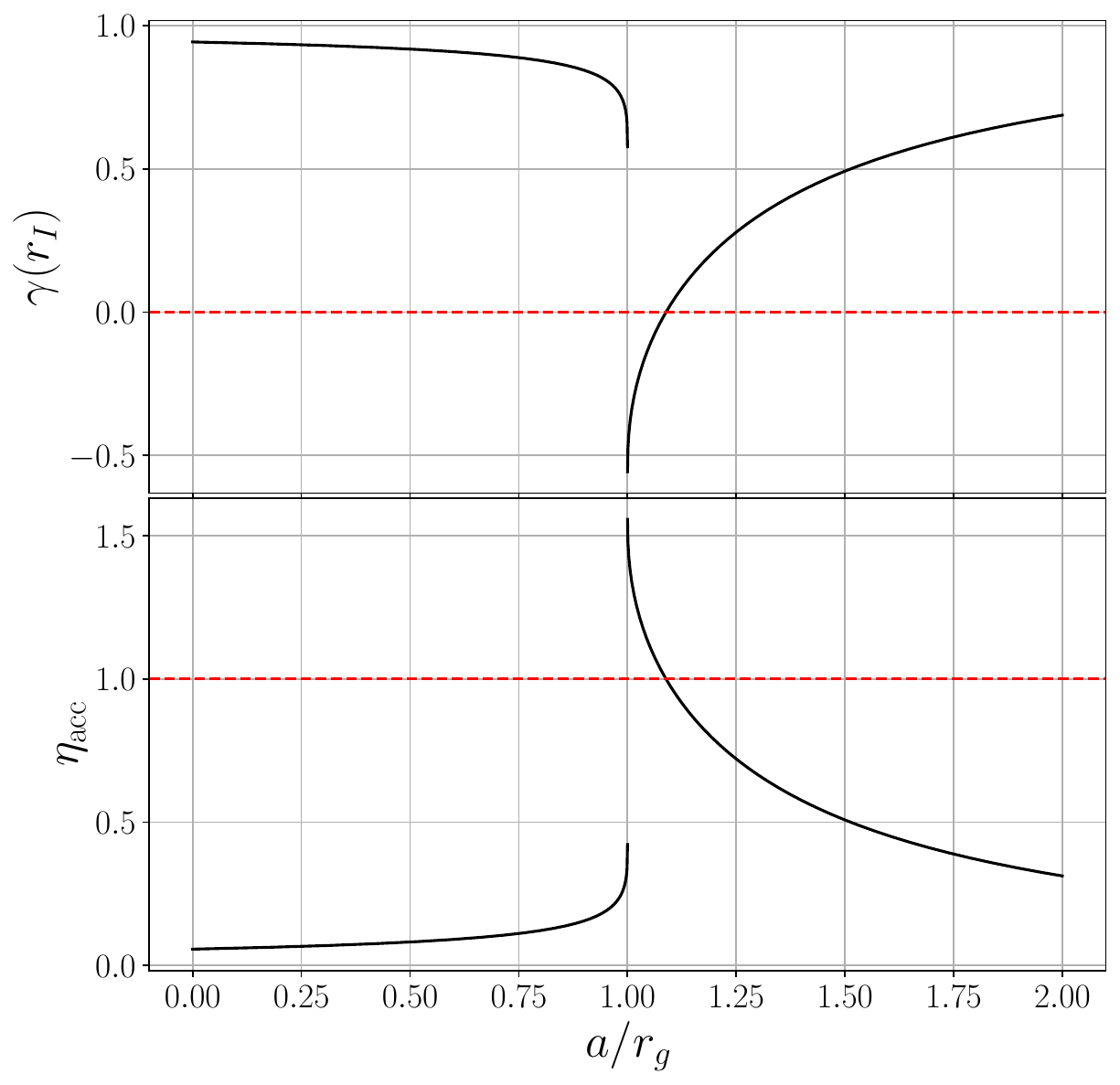}
\caption{{\it Upper:} The dimensionless energy of the ISCO circular orbit. {\it Lower:} The efficiency of the thin disc accretion  process, both plotted as a function of Kerr metric spin parameter. Both parameters are discontinuous at $a_\star = 1$.  }
\label{fig2}
\end{figure}

The properties of $\gamma(r_I)$ are displayed in Fig. \ref{fig2}. The minimum energy of an ISCO circular orbit is\footnote{We use the notation $1^\pm \equiv \lim_{\epsilon \to 0} 1 \pm \epsilon$, where $\epsilon > 0$.  } 
\beq
 \gamma_{\rm min} = \gamma_I \left(1^+\right) = -{1\over\sqrt{3}},
\eeq
and the circular orbit at the critical spin $a_\star = 2^{5/2}/3^{3/2}$ has zero energy. This means that the maximum accretion efficiency is equal to 
\beq
\eta_{\rm max} = 1 - \gamma_{\rm min}  = 1 + {1\over\sqrt{3}} > 1. 
\eeq
The efficiency of accretion for some super-extremal Kerr metrics is greater than $100\%$ (Fig. \ref{fig2}). Spin energy of the {singularity} is being tapped directly by the accretion process. In fact, more than the entire rest mass of the fluid elements may be radiated away for spins in the range
\beq
1 < a/r_g < 2^{5/2}/3^{3/2} ,
\eeq 
where the final limit corresponds to the spacetime with minimum ISCO radius $r_I = 2r_g/3$, an orbit with zero energy. 

\subsection{Intra-ISCO inspiral} 
The final inspiralling trajectory of a test particle which has crossed the super-extremal ISCO can be computed by analysing the effective potential of the ISCO circular orbit. The effective potential is defined by the orbital equation 
\beq
\left(U^r \right)^2 + V_{\rm eff}(r) = 0 ,
\eeq
and can be derived from 
\beq
g_{\mu \nu }U^\mu U^\nu = -c^2.
\eeq
In other words 
\beq
V_{\rm eff}(r) = {1\over g_{rr}} \left(c^2 + U_t U^t + U_\phi U^\phi\right).
\eeq
Writing $U^t = g^{t\mu}U_\mu$, $U^\phi = g^{\phi \mu} U_\mu$, using equations \ref{orbang} \& \ref{orben}, and recalling that $U_\phi$ and $U_t$ are conserved along geodesics  allows us to determine the effective potential for a given circular orbit radius. 

\begin{figure}
\includegraphics[width=0.45\textwidth]{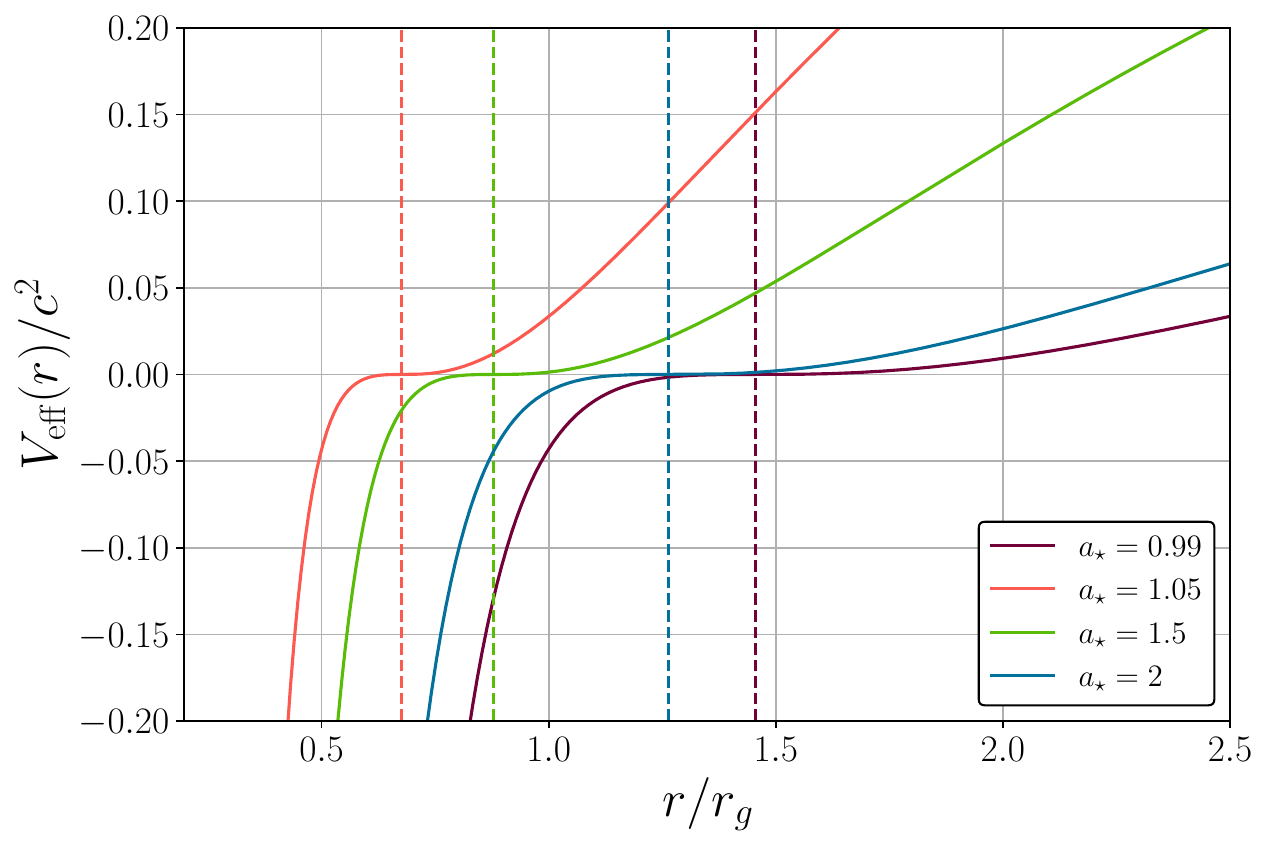}
\caption{The effective potential (defined in text) of circular orbits at the ISCO, plotted as a function of radius for different Kerr metric spin parameters (see Figure legend).  The ISCO radius is plotted as a vertical dashed line for each spin parameter. We see that super-extremal ISCOs are unstable to inward plunges.   }
\label{Pots}
\end{figure}

\begin{figure}
\includegraphics[width=0.45\textwidth]{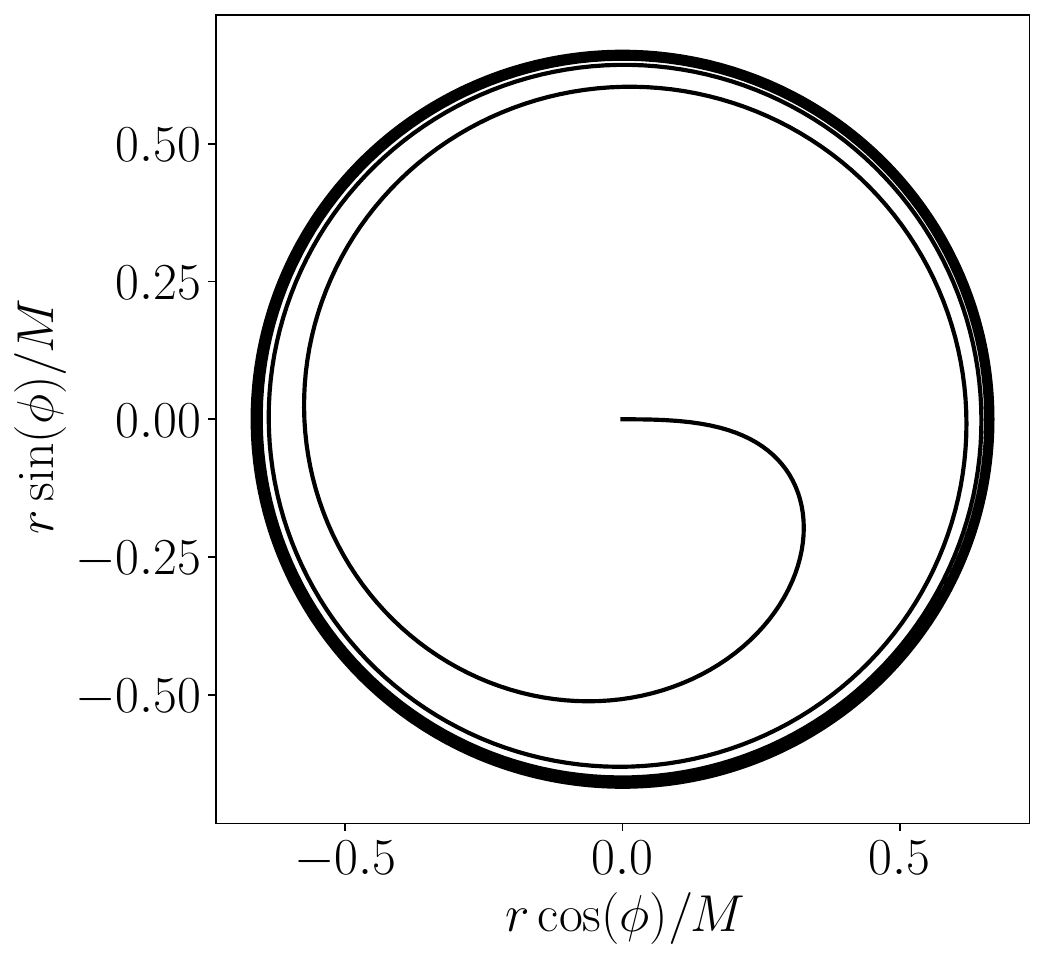}
\includegraphics[width=0.45\textwidth]{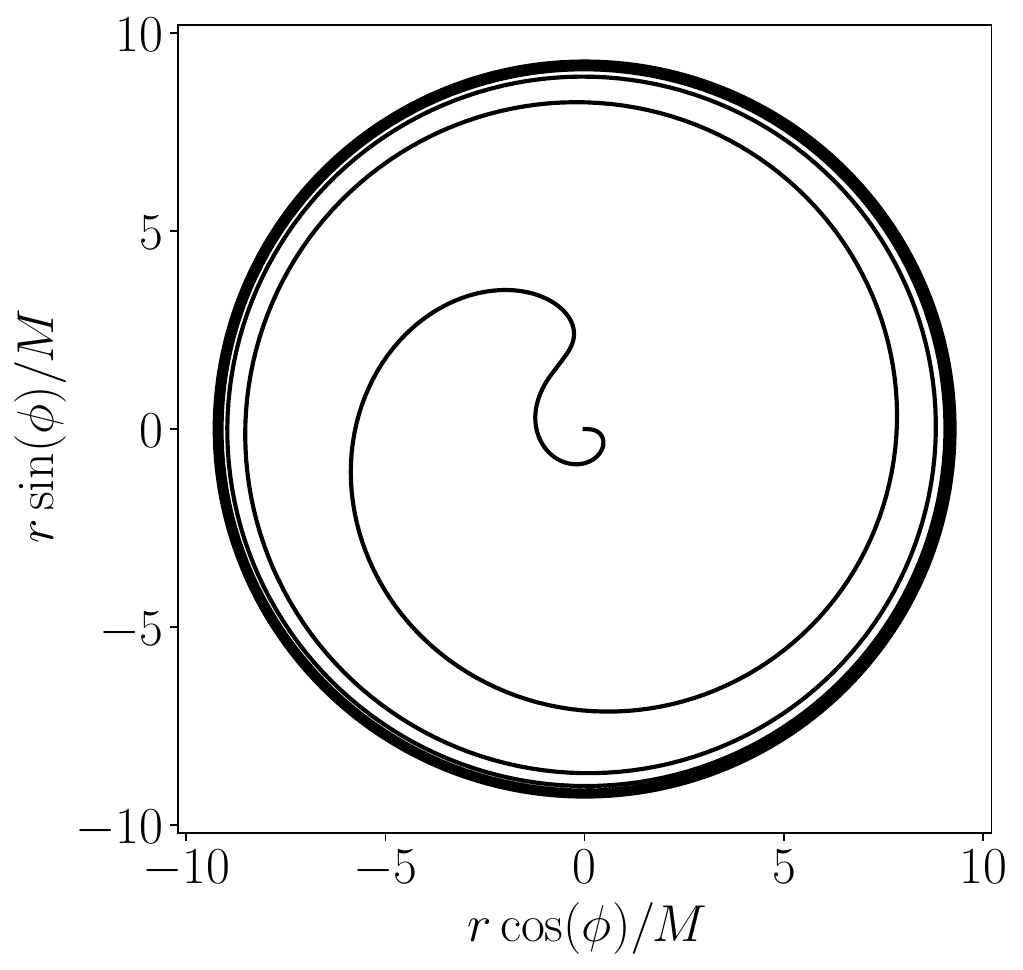}
\caption{Example inspirals from the ISCO to singularity of the super-extremal Kerr metric with dimensionless spins $a_\star = \pm1.1$ (top/bottom).  }
\label{Spiral}
\end{figure}

In Fig. \ref{Pots} it can be clearly seen that the ISCO radius represents an inflection point of the effective potential, for all Kerr spins. A perturbation inwards from the super-extremal Kerr singularity will pick up an increasing radial velocity. As the effective potential is an inflection point  with zero amplitude at the ISCO, the radial velocity within the ISCO is given by \citep[cf.][]{MummeryBalbus22PRL}
\beq
U^r = - c \sqrt{2r_g \over 3r_I} \left({r_I \over r} - 1\right)^{3/2} .
\eeq

This radial velocity, when combined with the intra-ISCO azimuthal velocity 
\beq
U^\phi =  - g^{\phi t} \gamma_I + g^{\phi\phi} J_I = {2r_g\gamma_I ac + J_I(r -2r_g) \over r(r^2 - 2r_gr + a^2)} ,
\eeq
can be integrated to determine the full inspiralling trajectory of test particle orbits in the super-extremal Kerr metric. The solutions of the resulting azimuthal integral describing these  trajectories is of an identical form to that found in \citep{MummeryBalbus22PRL}, to which we refer the reader for further details. Explicitly, the solution is 
\begin{multline}\label{phisol}
\phi(r) = C_0\sqrt{r \over r_I - r}  + C_- \tanh^{-1}\left(\sqrt{{r_- \over r}{ (r - r_I) \over (r_--r_I)}}\right) \\
- C_+ \tanh^{-1}\left(\sqrt{{r_+ \over r}{ (r - r_I) \over (r_+ -r_I)}}\right). 
\end{multline}
where we have defined the constants 
\begin{align}
C_0 &= \sqrt{6r_I\over r_g}\ {2r_g(J_I-ac\gamma_I) - r_IJ_I\over r^2_I -2r_gr_I +a^2} \label{c0} ,\\
C_\pm &=  {t_\pm\over t_+^2 - t_-^2}\sqrt{6r_I\over r_g }\ {2r_g(J_I-ac\gamma_I)(1+t_\pm^2) -r_IJ_It_\pm^2 \over r^2_I -2r_gr_I +a^2},  \label{cpm}
\end{align}
and the complex radii 
\beq
r_\pm / r_g = 1 \pm \sqrt{1 - a_\star^2} ,
\eeq
and 
\beq
t_\pm = \sqrt{x_\pm \over 1 - x_\pm}, \quad x_\pm \equiv r_\pm/r_I . 
\eeq
Note that the principal branch of the $\tanh^{-1}(z)$ function in the complex plane must be used. 
Example inspirals for $a_\star = \pm1.1$ metrics are shown in Fig. \ref{Spiral}.

\section{The evolution of a super-extremal Kerr singularity} 
A black hole which accretes matter over an extended period of time will evolve, with both its mass and angular momentum changing as a result of it accruing matter. The same is true for the naked singularities of the super-extremal Kerr metric. If material drops out of an accretion disc at the ISCO, and carries with it the ISCO circular orbit energy and angular momentum, then accreting a fluid element with rest mass  $\delta m_0$ will increase the angular momentum and energy of the singularity by \citep{Bardeen1970, Thorne1974} 
\beq
\delta J = U_\phi(r_I) \delta m_0, \quad \delta M = - U_t(r_I) \delta m_0 / c^2. 
\eeq
Evolution equations for the mass and angular momentum parameters of the Kerr metric can therefore be derived, and have the form 
\beq
{{\rm d} M \over {\rm d} m_0} = - U_t(r_I)/c^2 ,
\eeq
and 
\beq
{{\rm d} a_\star \over {\rm d} m_0} = {{\rm d} \over {\rm d} m_0} \left({Jc \over GM^2} \right) = {c \over GM^2} U_{\phi}(r_I) + {2 a_\star \over Mc^2} U_t(r_I) .
\eeq
It was \cite{deFelice78} who first showed that for super extremal Kerr singularities with spin parameters in the range $1 < a_\star < 2^{5/2}/3^{3/2}$ the negative-energy orbits of fluid elements at the ISCO  (discussed above) actually leads to the  spinning {\it down} of a super-extremal Kerr singularity which is accreting through a thin disc.  This disc spin-down result in fact holds for {\it all} super-extremal Kerr metrics, as we now demonstrate.   Using the general form of $U_\phi(r_I)$ and $U_t(r_I)$ valid for both sub- and super-extremal spins (derived above, see eqs.  \ref{isco_energy} and \ref{isco_ang_mom})
\begin{align}
U_\phi(r_I) &= 2\sqrt{3}r_g c \left( 1 - {2a\over 3\sqrt{r_gr_I} }\right) {\rm sgn}(\sqrt{r_I r_g}
- a)  ,\\
U_t(r_I) &= - {4c^2\over\sqrt{3}}\left(r_g\over r_I\right)^{1/2} \left( 1 -{3a\over 4\sqrt{r_gr_I}}\right){\rm sgn}(\sqrt{r_I r_g}
- a)  ,
\end{align}
the evolution equation for the dimensionless spin parameter can be manipulated to 
\beq\label{a_evol_def}
M {{\rm d} a_\star \over {\rm d} M} = M  {{\rm d  }a_\star  \over {\rm d} m_0} {{\rm d} m_0 \over {\rm d} M}  =  - {c^3 \over GM} { U_{\phi}(r_I) \over U_t(r_I)} - {2 a_\star} ,
\eeq
which equals 
\beq
M {{\rm d} a_\star \over {\rm d} M} = {3 \over 2 } \sqrt{r_I \over r_g} \left({ 1 - {2a_\star \over 3} \sqrt{r_g \over r_I} \over 1 - {3 a_\star \over 4} \sqrt{r_g \over r_I}} \right)- 2a_\star ,
\eeq
or equivalently 
\beq
M {{\rm d} a_\star \over {\rm d} M} = {{3\over 2} \sqrt{r_I \over r_g} - 3 a_\star + {3a_\star^2 \over 2 } \sqrt{r_g \over r_I} \over 1  - {3 a_\star \over 4} \sqrt{r_g \over r_I}} .
\eeq
We denote the dimensionless ISCO radius as $x_I\equiv r_I/r_g$, which is a function only of dimensionless black hole spin and is defined by 
\beq
x_I^2  -    6x_I + 8a_\star\sqrt{x_I} - 3a_\star^2  = 0. 
\eeq
Differentiating the above expression with respect to  $x_I$ leads to 
\beq
x_I {{\rm d}a_\star \over {\rm d} x_I} = -{1 \over 8}\sqrt{1 \over x_I} \left({2 x_I^2 - 6 {x_I} + 4 a_\star \sqrt{x_I} \over 1 - {3 a_\star \over 4} \sqrt{1 \over x_I}} \right) .
\eeq
A final factorisation of the spin evolution equation
\beq
M {{\rm d} a_\star \over {\rm d} M} = {1\over 4}\sqrt{1 \over x_I} \left( {6 {x_I}   - 12 a_\star \sqrt{x_I}  + 6 a_\star^2  \over 1  - {3 a_\star \over 4} \sqrt{1\over x_I}} \right), 
\eeq
before substituting for $a_\star^2$ from the ISCO definition, leads to the dramatic simplification 
\beq
M {{\rm d} a_\star \over {\rm d} M} = - 2 x_I {{\rm d} a_\star \over {\rm d} x_I}. 
\eeq
There are two solutions to this set of equations, the first is that ${{\rm d} a_\star / {\rm d}M} \neq 0$, and as such
\beq
{{\rm d} \ln M \over {\rm d} \ln x_I} = - {1\over 2} ,
\eeq
and therefore 
\beq
x_I M^2 = {\rm const} \quad \Rightarrow \quad r_I M = {\rm const} .
\eeq
This result was first derived by \citet{Bardeen1970} in the context of sub-extremal Kerr black holes, but in fact holds for all solutions of the general Kerr metric.  

Note that there is an additional solution to this evolutionary equation, one in which both ${{\rm d} a_\star / {\rm d}M} = 0$ and ${{\rm d} a_\star / {\rm d}x_I} = 0$. This solution is valid for $a_\star = x_I = 1$. We therefore find that the extremal prograde spin $a_\star = 1$ is the limiting value obtained via accretion {\it for all initial values of the Kerr metric spin parameter}. 

From this result we may solve for $M(m_0)$ and $a_\star(m_0)$. The mass evolution equation takes the form 
\beq
{{\rm d} r_g \over {\rm d} m_0} = {s G \over c^2} \sqrt{1 - {2 r_g \over 3r_I}} = {s G \over c^2}  \sqrt{1 - {2 r_g^2 \over 3r_{I, i}r_{g, i}}},
\eeq
where 
\beq
s = \begin{cases}
+1 , \quad a_\star \leq 1, \\
\\
-1 , \quad 1 < a_\star \leq 2^{5/2}/3^{3/2}, \\
\\
+1 , \quad a_\star > 2^{5/2}/3^{3/2}. 
\end{cases}
\eeq
Integrating,  we find
\beq
{GM(m_0) \over c^2}  = \sqrt{3 r_{g, i} r_{I, i} \over 2} \sin\left( C + s {\sqrt{2} Gm_0 \over \sqrt{{3r_{g, i} r_{I, i}}} c^2}  \right),
\eeq
where 
\beq
C =\sin^{-1} \left({\sqrt{2} GM_i \over \sqrt{3 r_{g, i} r_{I, i}} c^2 }\right) .
\eeq
Then 
\beq
r_I(m_0) = r_{I, i} \left({M_i \over M(m_0)}\right),
\eeq
and 
\begin{equation}
 a_\star = {4 \over 3} \sqrt{r_I \over r_g} \pm \sqrt{{1\over 3}\left({r_I \over r_g}\right)^2 - {2\over 9} \left({r_I \over r_g}\right)} .
\end{equation}
These equations are all valid provided $a_\star \neq 1$. 

\begin{figure}
\includegraphics[width=0.5\textwidth]{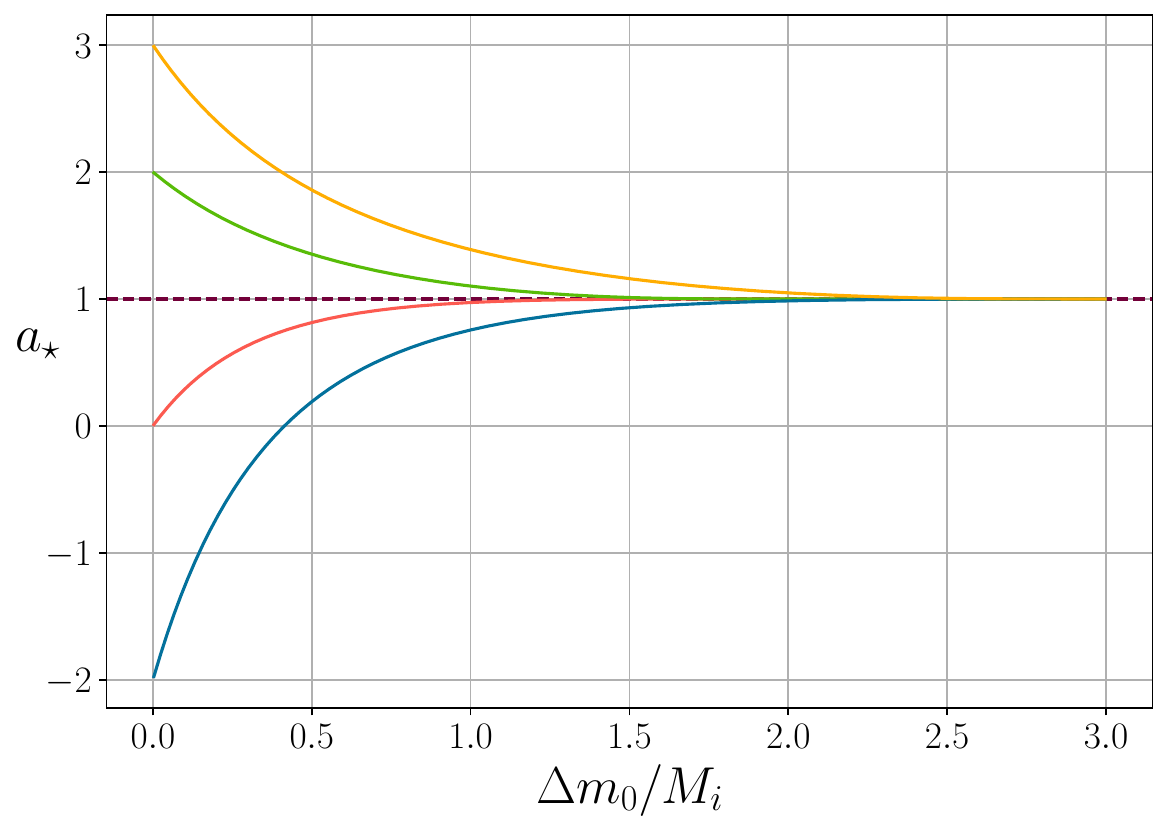}
\caption{The evolution of the Kerr spin parameter $a_\star$ as a function of accreted rest mass $\Delta m_0$ (plotted in units of the initial metric parameter $M_i$). The dimensionless spin $a_\star = 1$ is a global attractive point for all initial spins. }
\label{SpinEvol}
\end{figure}

In Fig. \ref{SpinEvol} we demonstrate that the dimensionless spin $a_\star = 1$ is a global attractive point for all initial spins. Accretion of a prograde disc spins {\it down} a super-extremal Kerr solution to that of an extremal solution. In this sense super-extremal Kerr naked singularities are formally unstable, and will in finite time revert to a Kerr black hole solution with $a_\star = 1$ through the accretion of surrounding material. 

{There are two additional refinements to the general evolutionary picture we have discussed here which must be taken into account. Firstly,}  \citet{Thorne1974} demonstrated that for Kerr black holes which are spin-up via accretion, the limiting value of the spin parameter is in fact $a_\star = 0.998$, not $a_\star = 1$, due to the preferential  capture of photons emitted from the accretion flow with angular momentum retrograde with respect to the black hole's spin axis. There is no such Thorne limit for discs around super-extremal singularities, as all photons emitted with angular momentum in the prograde direction escape to infinity \citep[e.g.,][a result discussed further below]{Chandrasekhar83}, and only retrograde photons hit $r=0$, further spinning down the singulairty towards $a_\star = 1$. 

{Secondly, it is likely not the case that fluid elements drop out of the disc at the ISCO and carry with them precisely their ISCO angular momentum and energy. Indeed, GRMHD simulations generally find horizon angular momenta which differ from their ISCO values by of order $\sim$ a few percent \citep[e.g.,][]{Shafee08, Noble10, Schnittman16}, with magnetic stresses transferring some angular momentum from the plunging material back to the main body of the disc. This reduces the angular momentum flux onto the singularity, and will in general make spin up less efficient.  

While the precise modifications to the results presented here will depend on the details of the angular momentum transport by magnetic stresses, it is possible to produce an estimate for the reduced rate of spin parameter change from magnetic stresses. Let us assume that the fluid's singularity angular momentum (equal to the horizon angular momentum in the black hole case)  is reduced by a factor $\Delta J$, i.e.,  }
\begin{equation}
    J_{\rm singulairty} = J_I - \Delta J . 
\end{equation}
{Kerr metric circular orbits follow the energy-angular momentum relation \citep[e.g.,][]{PageThorne74} }
\begin{equation}
    U_t' = - \Omega U_\phi ' , \quad \Omega = {\sqrt{GM/r^3} \over 1 + a\sqrt{r_g/r^3}} ,
\end{equation}
{where a prime denotes a radial gradient. While the fluid motion will not be that of circular motion, we can use this result to estimate the fluid's (dimensionless) energy parameter at the singularity to be}
\begin{equation}
    \gamma_{\rm singularity} = -U_t(r_I)/c^2  - \Omega_I \Delta J/c^2, 
\end{equation}
{where $\Omega_I$ is the angular velocity of the disc material at the ISCO. Equation \ref{a_evol_def} therefore becomes }
\beq
 {{\rm d} a_\star \over {\rm d} \ln M}  =  {c \over GM} { J_I \over \gamma_I}\left[ {1 - \Delta J/J_I \over 1 - \Omega_I \Delta J / \gamma_I c^2 } \right] - {2 a_\star} .
\eeq
{The term in square brackets is the leading order correction for small $\Delta J$. Taylor expanding for small $\Delta J$, and simplifying with the 4-velocity normalisation condition $U^\phi_I J_I - U^t_I\gamma_I c^2 = -c^2$, it is clear to see that this correction reduces the rate at which black holes are spun up by a given change in mass parameter $M$: }
\beq
 {{\rm d} a_\star \over {\rm d} \ln M}  \simeq  {c \over GM} { J_I \over \gamma_I}\left[ 1 - {\Delta J \over J_I \gamma_I U^t_I } \right] - {2 a_\star} .
\eeq
\section{The super-extremal disc temperature }
The temperature profile of a thin accretion flow fed by a constant mass flux $\dot M$ in the general Kerr metric $g^{\mu \nu}$ is given by \citep{NovikovThorne73, PageThorne74, Balbus17} 
\beq\label{temp_def}
\sigma T^4 = - {1 \over 2\sqrt{|g|}} \left( U^t \right)^2 {{\rm d} \Omega  \over  {\rm d} r} \left[ {\dot M \over 2\pi} \int^r {U_\phi'  \over U^t } \, {\rm d}r + {{\cal F}_{\cal J} \over 2\pi} \right] ,
\eeq
where 
\beq
\Omega \equiv {U^{\phi} \over U^t} ,
\eeq
$g$ is the metric determinant, and $\dot M$ is the constant mass flux through the disc, while ${\cal F}_{\cal J}$ is an integration constant relating to the angular momentum flux through the disc. The angular momentum gradient $U_\phi'$ is given by eq. \ref{grad_j}, while $U^t(r, a)$ is given in the stable disc region by 
\beq
U^t(r, a) = {1 + a\sqrt{r_g/r^3} \over \left(1 - 3r_g/r + 2a\sqrt{r_g/r^3}\right)^{1/2}}.
\eeq
The key result therefore is to evaluate the integral 
\begin{align}
{\cal I}_a &\equiv {\dot M \over 2\pi} \int^r {U_\phi'  \over U^t } \, {\rm d}r , \nonumber  \\ 
&= {\dot M  c \over 4\pi} \int^r \sqrt{r_g \over r} {r^2 - 6r_gr - 3a^2 + 8a\sqrt{r_g r} \over r^2 - 3r_gr + 2a\sqrt{r_gr}}\, {\rm d}r .
\end{align}
With $x \equiv \sqrt{r/r_g}$ and $a_\star = a/r_g$, we have
\beq
{\cal I}_a = {\dot M r_g c \over 2 \pi } \int^{\sqrt{r/r_g}}  {x^4 - 6x^2 - 3a_\star^2 + 8a_\star x \over x^4 - 3x^2 + 2a_\star x }\, {\rm d}x ,
\eeq
which can be solved with standard partial fraction techniques. The approach is to re-write the lower cubic in terms of its roots
\beq\label{roots}
x^3 - 3x + 2a_\star = (x - x_1)(x-x_2)(x-x_3) ,
\eeq
 before expanding the resultant fraction and integrating term by term. The roots of the cubic, using standard  trigonometric identities, are given in \citet{PageThorne74} as 
\beq
x_\lambda = 2 \cos\left[ {1\over 3} \cos^{-1}\left(-{a \over r_g}\right) - {2\pi\lambda\over3}\right] ,
\eeq
with $\lambda = 0, 1, 2$. However, it is clear to see that for super-extremal spins $|a| > r_g$ this standard approach will run into issues, as the inverse cosine of a number greater than 1 is complex. The integral itself is purely real, and care must be taken in resolving this apparent issue. We solve this integral in a manifestly real manner in Appendix \ref{integral_app}. {We shall assume that the disc fluid extends down to the ISCO, appropriate for comparison to accreting sources in the so-called ``soft'' state.} The solution has a fundamentally different form to the classic Page and Thorne solution, a result of the complex nature of two of the roots of the cubic. In full we have 
\begin{multline}
{\cal I}_a = {\dot M r_g c \over 2 \pi } \Bigg[ x - {3a_\star \over 2} \ln(x) + A \ln (x - \xi)  \\ + B  \ln\left({\left(x + {1\over 2 }\xi \right)^2 + \psi^2 }\right) \\ + C \tan^{-1} \left({\psi \over x + {1\over 2}\xi + \sqrt{\left(x + {1\over 2 }\xi\right)^2 + \psi^2} }\right) \Bigg],
\end{multline}
where $A, B, C, \xi$ and $\psi$ are all real valued explicit functions of the dimensionless black hole spin $a_\star$: 
\begin{align}
\xi &= -2 \cosh\left[ {1\over 3} \cosh^{-1} \left({|a_\star|}\right)\right] , \\
\psi &= \sqrt{3 \cosh^2\left[ {1\over 3} \cosh^{-1} \left({|a_\star| }\right)\right] - 3} , \\
\theta &= 3 - 2\cosh^2\left[ {1\over 3} \cosh^{-1} \left({|a_\star|}\right)\right] , \\
A &= {2 \xi - a_\star(1 + \xi^2)  \over 2(1 - \xi^2)} , \\
B &= {2 \xi \psi^2 (1 + {1\over 2} \xi a_\star) - \xi - a_\star \left(1 -\theta^2 \right) \over 2\left( \left( 1 -\theta\right)^2 + \xi^2 \psi^2\right)} ,  \\
C &=   {2\xi \psi \left(\xi + a_\star  \left(1 + \theta  \right) \right) + 4 \psi (1 + {1\over 2} a_\star \xi) \left(1 - \theta  \right) \over \left( \left( 1 -\theta\right)^2 + \xi^2 \psi^2\right)} .
\end{align}
We can therefore write the solution for the disc temperature profile in closed form 
\begin{multline}\label{temperature}
\sigma T^4 = {3 G M \dot M \over 8 \pi r^3} \left({1 \over  1 - {3 / x^2} + {2 a_\star / x^3} }\right) \\ \Bigg[ 1 - {3a_\star \over 2x} \ln(x) + {A\over x} \ln (x - \xi)   + {B\over x}  \ln\left({\left(x + {1\over 2 }\xi \right)^2 + \psi^2 }\right) \\ + {C\over x} \tan^{-1} \left({\psi \over x + {1\over 2}\xi + \sqrt{\left(x + {1\over 2 }\xi\right)^2 + \psi^2} }\right) + {{\cal F}_{\cal J} \over \dot M\sqrt{r_g} x c } \Bigg]  ,
\end{multline}
where the final term in the square bracket represents the boundary condition of the location of ``vanishing stress'' in the flow. Note that some care is required for the precise limiting values $a_\star = \pm 1$, we discuss this further, and derive the temperature profiles at these limits, in section \ref{extremal_temp_lims}. 
\begin{figure}
\includegraphics[width=0.5\textwidth]{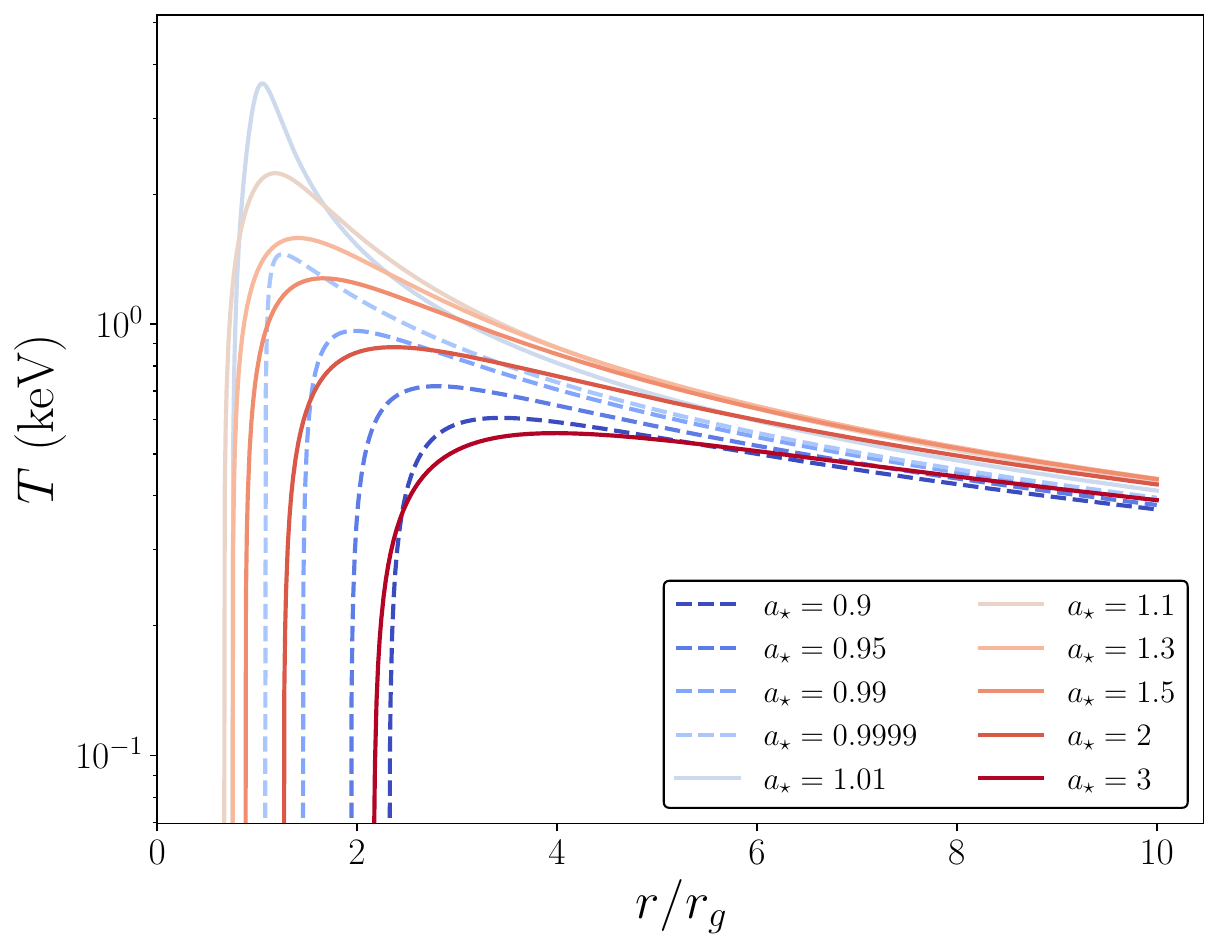}
\caption{Thin disc temperature profiles of vanishing ISCO stress discs evolving in the general Kerr metric. Dashed curves represent sub-extremal metrics, while solid curves represent super-extremal metrics. The other system parameters are $M = 10 M_\odot$ and $\dot M = 10^{15}$ kg/s.}
\label{TemperatureProfiles}
\end{figure}

Taking the location of vanishing stress to be the ISCO, as is often assumed for standard black hole discs, we generate disc temperature profiles for a number of dimensionless black hole spins $a_\star$ in Fig. \ref{TemperatureProfiles}. The dashed curves are sub-extremal metrics, while the solid curves are super-extremal solutions. 

\subsection{Gravitational red-shifts near to $a_\star = 1$ } 
Any {Kerr metric with} dimensionless spin satisfying  $|a_\star| > 1$ {has no} event horizon. The event horizon plays an important role in the observations of accreting systems, as it is the point at which the gravitational red-shift of any observed photons diverges. It at first appears therefore that there could be a discontinuity in the observed properties of any disc emission as the Kerr spin parameter changes from $a_\star = 1 - \epsilon$ to $a_\star = 1 + \epsilon$, where $\epsilon \ll 1$. 
\begin{figure}
\includegraphics[width=0.5\textwidth]{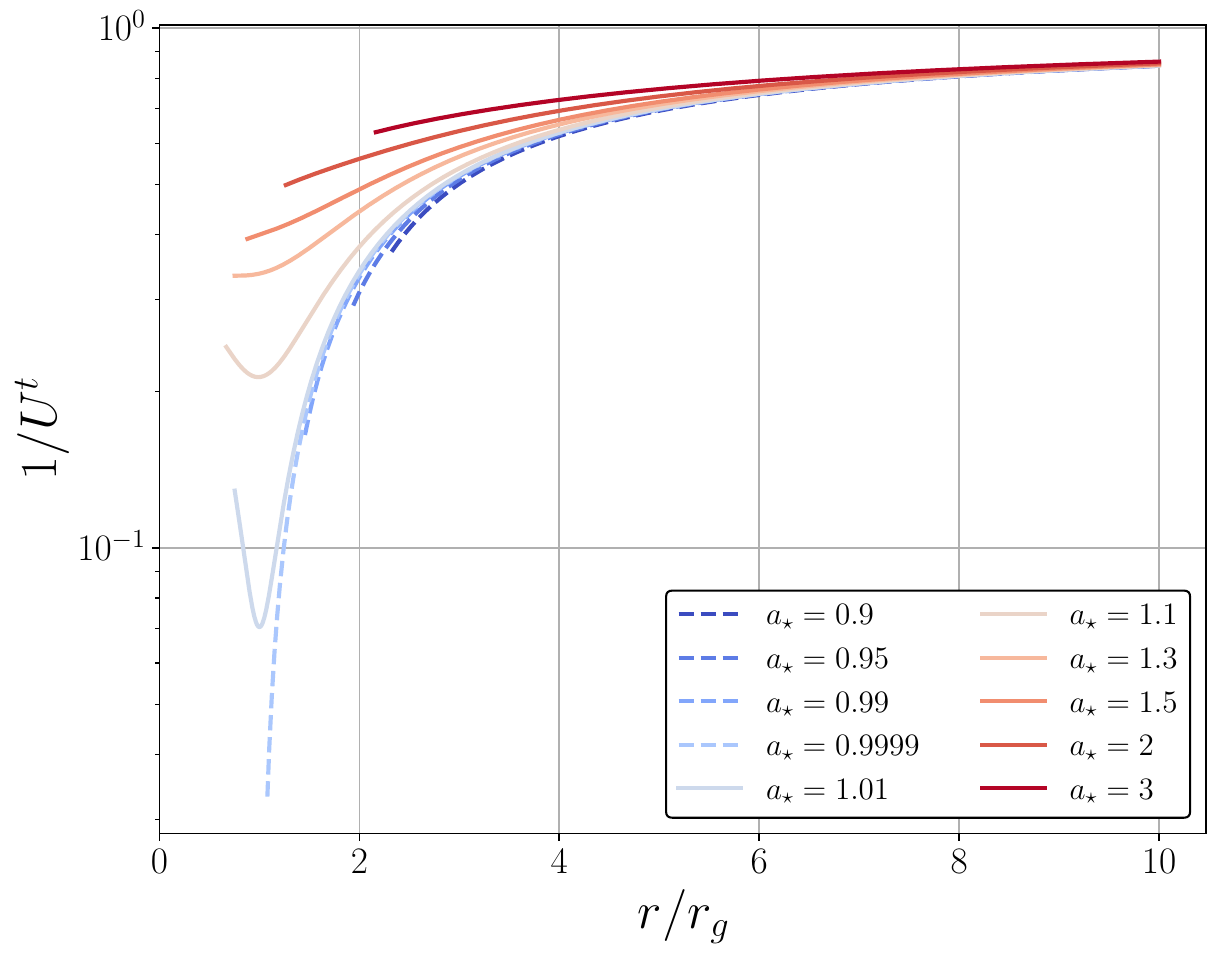}
\caption{The gravitational red-shifts of photons emitted from a disc radius $r$ in the general Kerr metric. Dashed curves represent sub-extremal metrics, while solid curves represent super-extremal metrics. Despite the event horizon {not existing} for $|a_\star| > 1$, the gravitational red shift is continuous in its properties as the Kerr spin parameter is changed through unity.  }
\label{GravitationalShifts}
\end{figure}

However, it transpires that this is not the case, and the transition from sub- to super-extremal gravitational red-shifts is broadly continuous. A photon observed at a large distance from the center of a Kerr metric, and which was emitted at a location $(r, \phi, \theta=\pi/2)$ in the disc has an energy shift given by \citep[e.g.,][]{MTW} 
\beq
{E_{\rm obs} \over E_{\rm emit}} = {1 \over U^t} \left(1 + {p_\phi \over p_t} \Omega \right)^{-1} . 
\eeq
In this expression $p_\phi$ and $-p_t$ are the angular momentum and energy of the emitted photon. These quantities are of course conserved along geodesics in the Kerr metric. All of the photons which reach the observer from a disc which is observed face-on have $p_\phi = 0$, and so the purely gravitational  part of the photon energy shift can be identified with $1/U^t(r)$. In Fig. \ref{GravitationalShifts} we plot the gravitational red-shift factor as a function of disc radius (i.e., radii external to the ISCO) for both sub- and super-extremal metrics. Despite the {non-existence of}  event horizon{s}  for $|a_\star| > 1$, there are still substantial gravitational red-shifts for Kerr metrics with $a_\star = 1 + \epsilon$. This ceases to be true for super-extremal metrics with spins $|a_\star| \gg 1$. 

\begin{figure}
\includegraphics[width=0.5\textwidth]{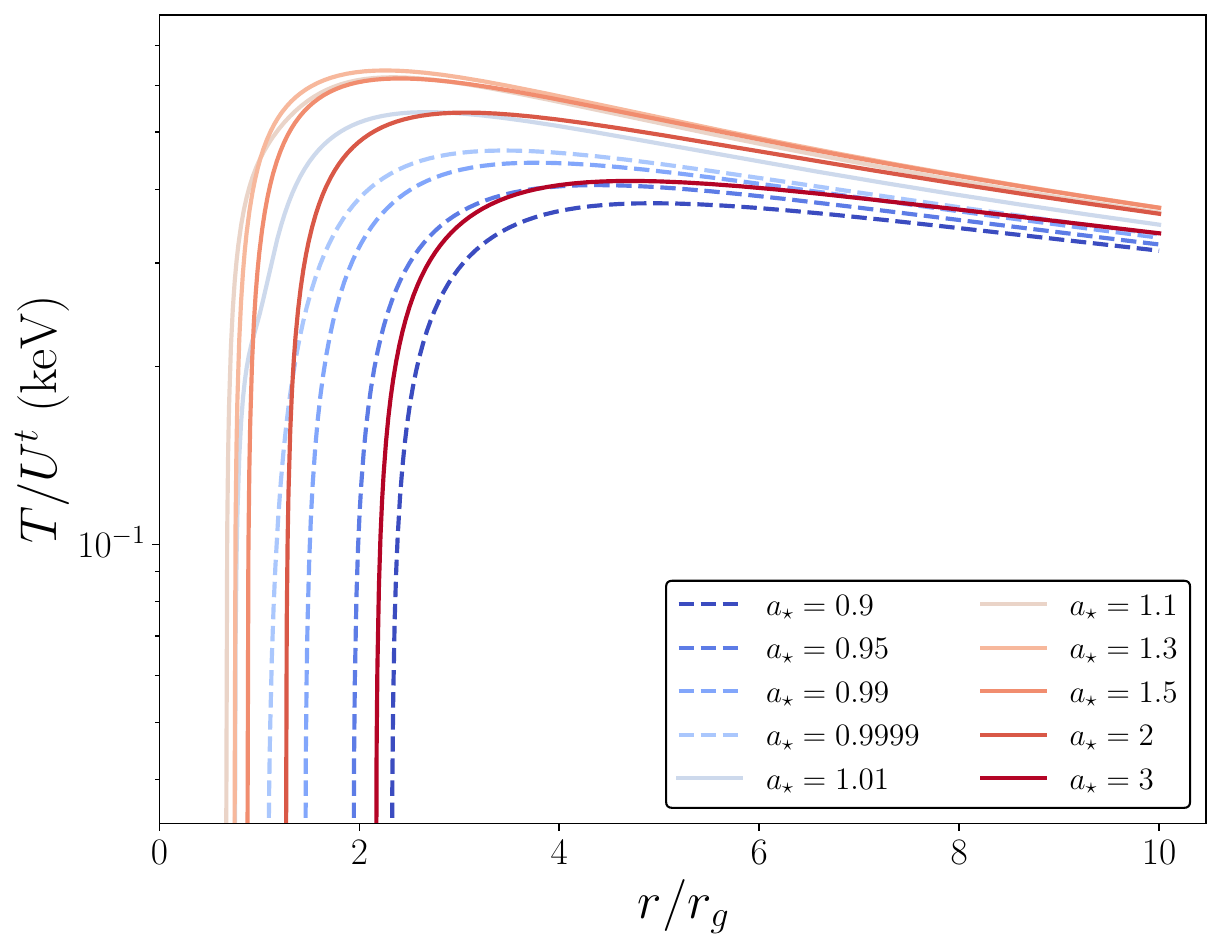}
\caption{The gravitationally red-shifted temperature profiles of the disc profiles plotted in Fig. \ref{TemperatureProfiles} (defined as $T(r)/U^t(r)$), as a function of disc radius $r$ for different values for the Kerr spin parameter $a_\star$. This parameter is a proxy for the temperature profile which would actually be observed from a face-on disc. }
\label{GravitationalTemperatureProfiles}
\end{figure}

The gravitationally red-shifted temperature profiles of these discs, $T(r)/U^t(r)$, which is a proxy for the temperature profile which would actually be observed from a face-on disc, is plotted in Fig. \ref{GravitationalTemperatureProfiles}. The evolution from sub- to super-extremal metrics is rather smooth. 

\section{Ray tracing and photon orbits }
To determine the observed properties of a super-extremal Kerr disc, one must follow the photon orbits through its spacetime. The orbital equations for the photon propagation are identical to that of the normal Kerr metric. We discuss our ray-tracing algorithm in Appendix \ref{ART}. The key difference between the two spacetimes is the lack of an event horizon surrounding a  super-extremal Kerr singularity. 

\begin{figure}
\includegraphics[width=0.49\textwidth]{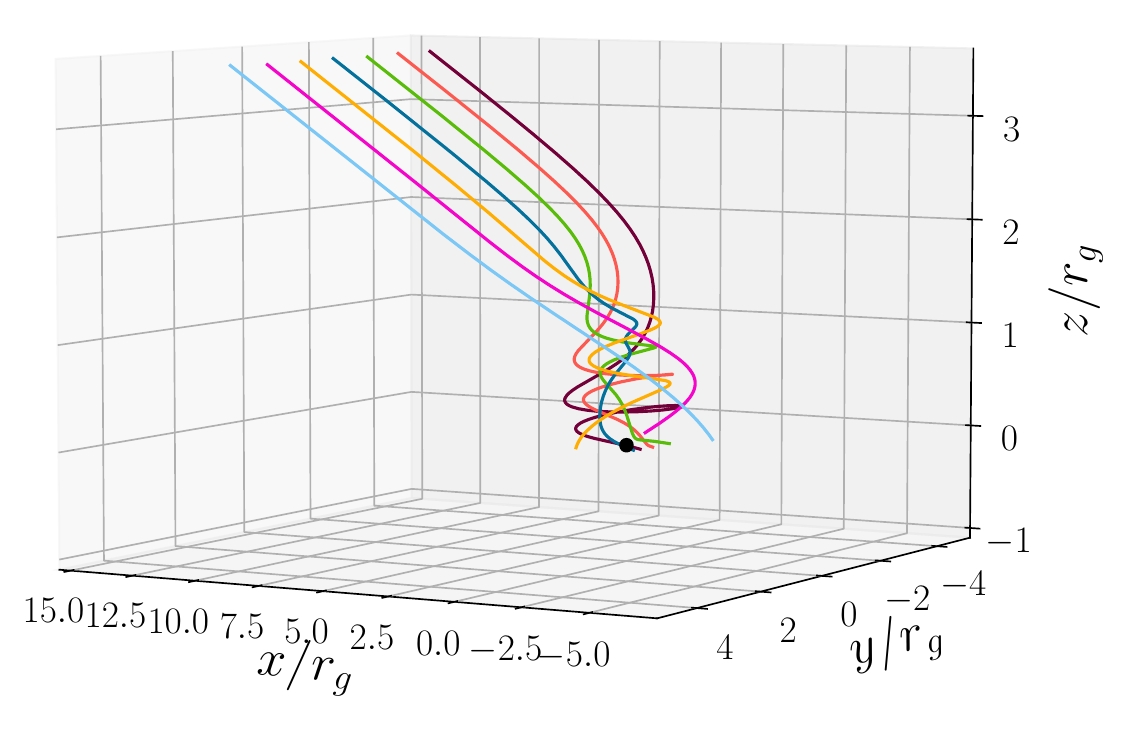}
\includegraphics[width=0.4\textwidth]{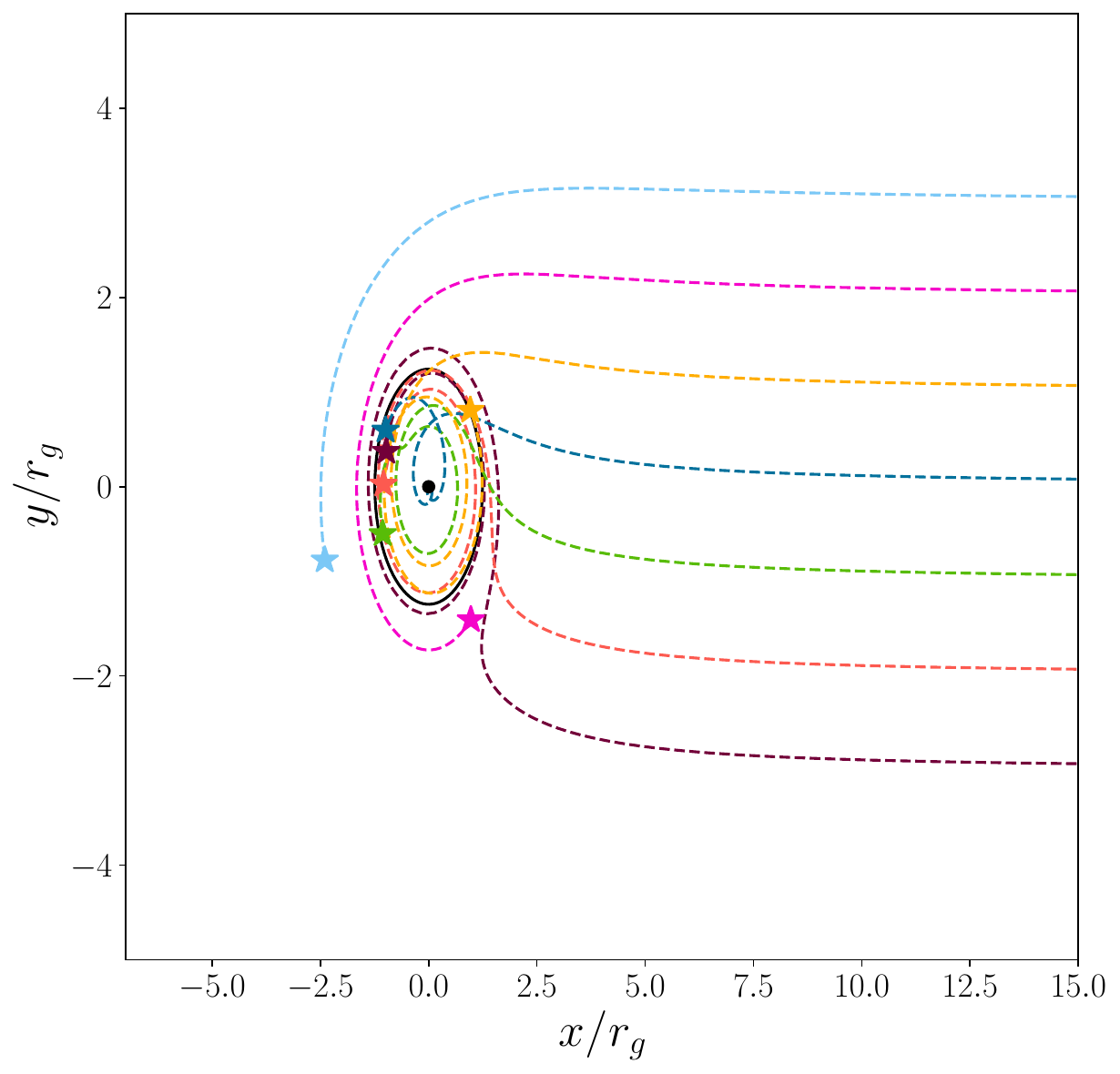}
\includegraphics[width=0.4\textwidth]{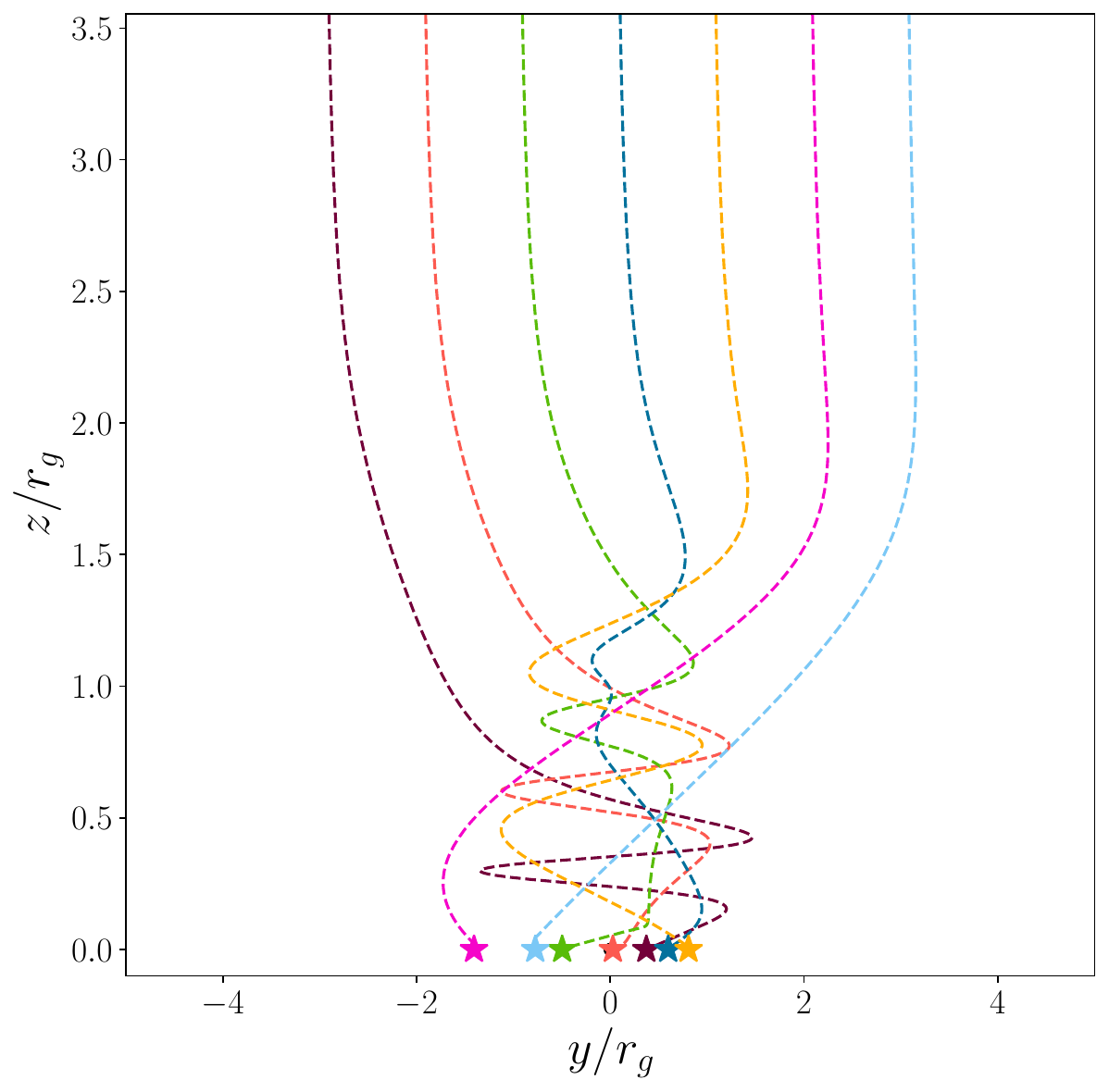}
\caption{Upper: Example trajectories for photon's propagating through the spacetime of a Kerr singularity with $a_\star = 1.03$. All of these photons escape towards a distant observer who is inclined at an angle of $80^\circ$ from the singularities spin axis. Some of the trajectories pass closer to the singularity than one gravitational radius, and so would be captured by the event horizon of a ``normal'' black hole solution (e.g., the yellow trajectory). The singularity is displayed as a black dot for aid in visualising the spacetime, but in reality has zero size. Middle/Lower: The trajectories projected into the $x-y$/$y-z$ planes, with stars denoting the location at which the trajectory intersects the $\theta = \pi/2$ disc plane.}
\label{RT1}
\end{figure}

In Fig. \ref{RT1} we show example trajectories for photon's propagating through the spacetime of a Kerr singularity with $a_\star = 1.03$. All of these photons escape towards a distant observer who is inclined at an angle of $80^\circ$ from the singularity's spin axis. Some of the trajectories displayed in Fig. \ref{RT1} pass closer to the singularity than one gravitational radius, and so would be captured by the event horizon of a ``normal'' black hole solution (e.g., the yellow trajectory). However, these photons may now continue to propagate and reach the observer in this super-extremal spacetime. The two lower panels of Fig. \ref{RT1} show $x-y$ and $y-z$ projections of the photon trajectories\footnote{We use quasi-Cartesian definitions of $x, y$ and $z$, related to Kerr metric $(r, \theta, \phi)$ by $x \equiv r \sin \theta \cos \phi, y \equiv r \sin\theta \sin \phi, z \equiv r \cos \theta$}. 

\begin{figure*}
\includegraphics[width=0.49\textwidth]{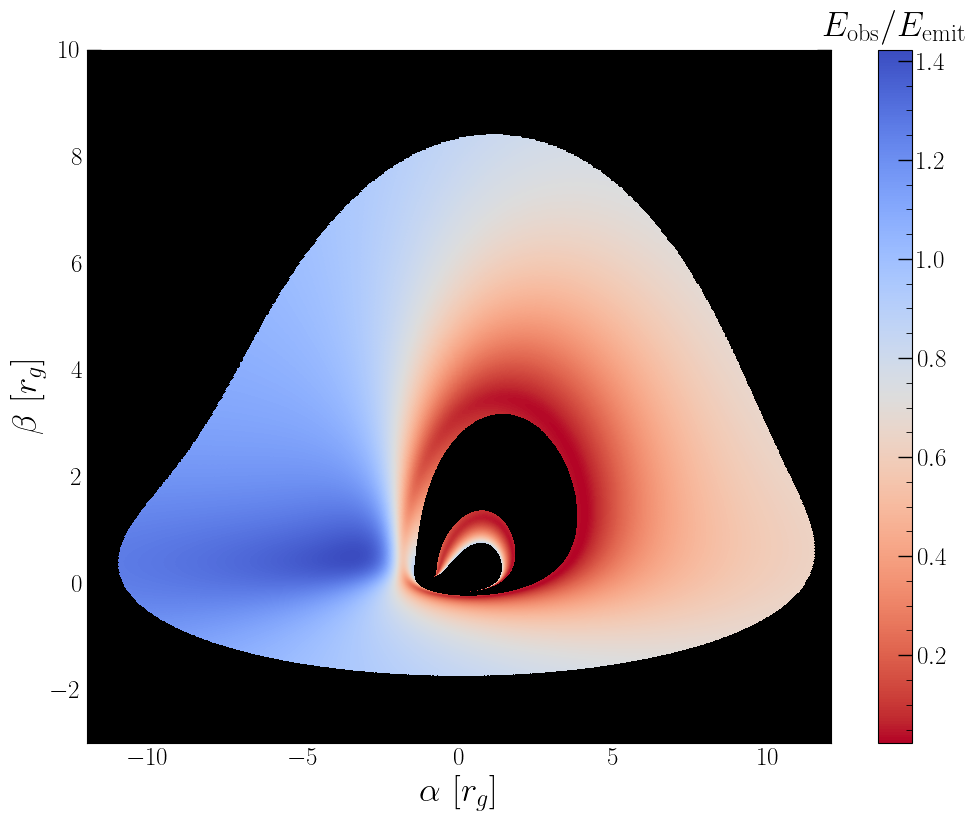}
\includegraphics[width=0.49\textwidth]{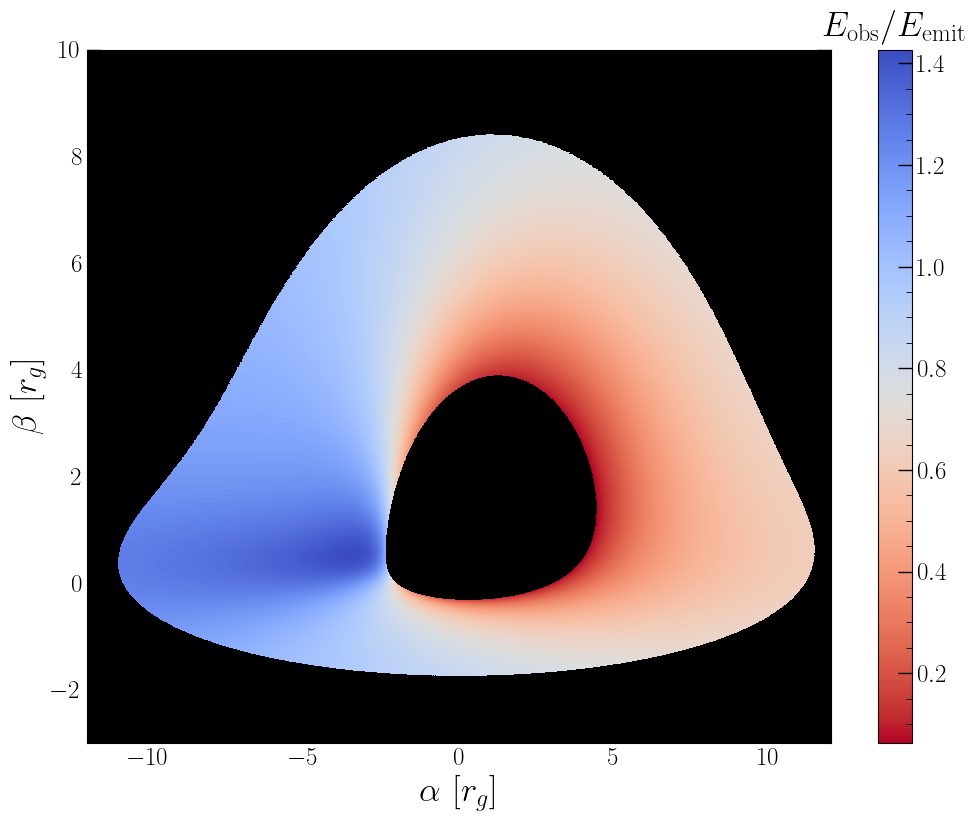}
\includegraphics[width=0.48\textwidth]{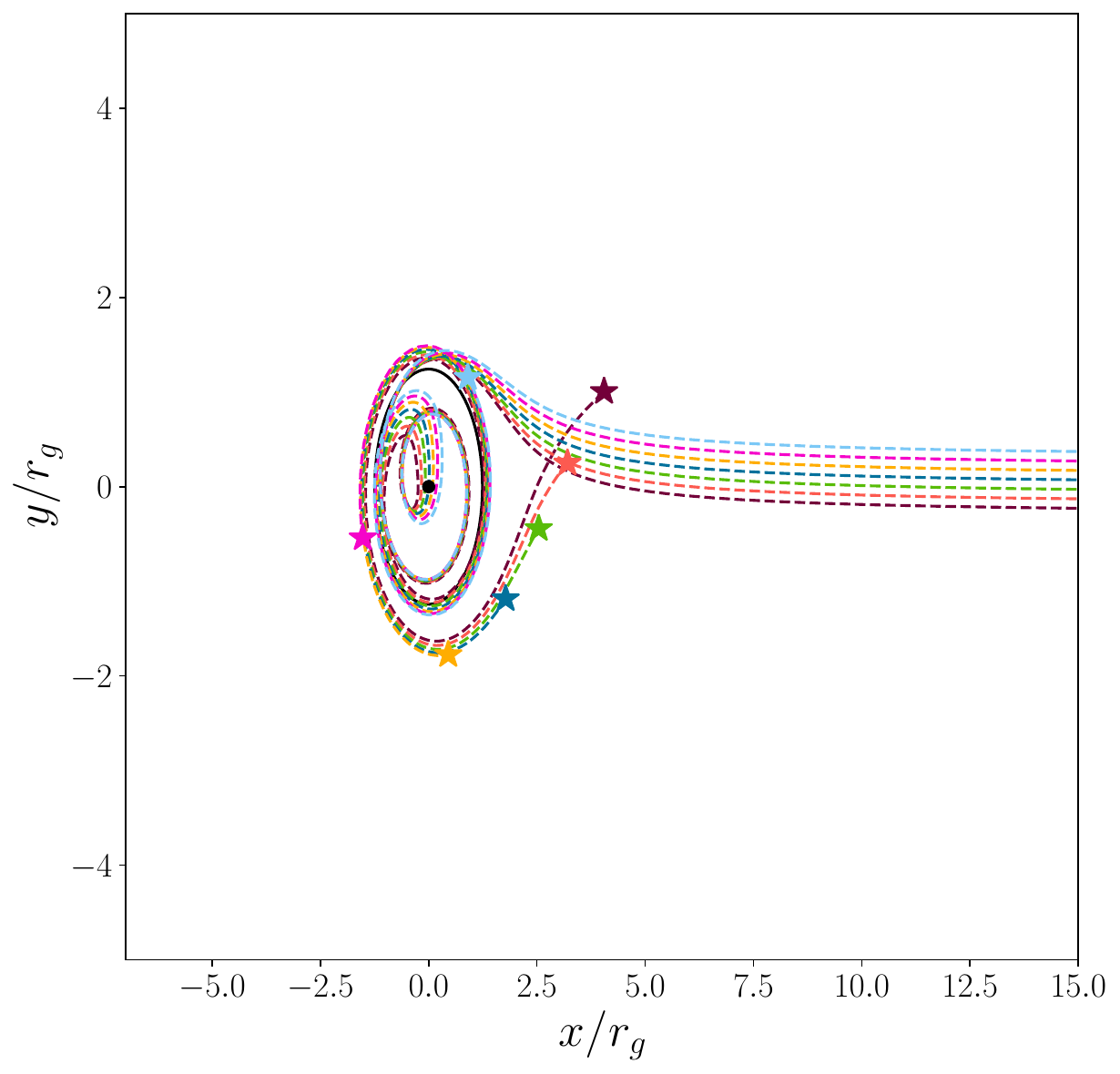}
\includegraphics[width=0.48\textwidth]{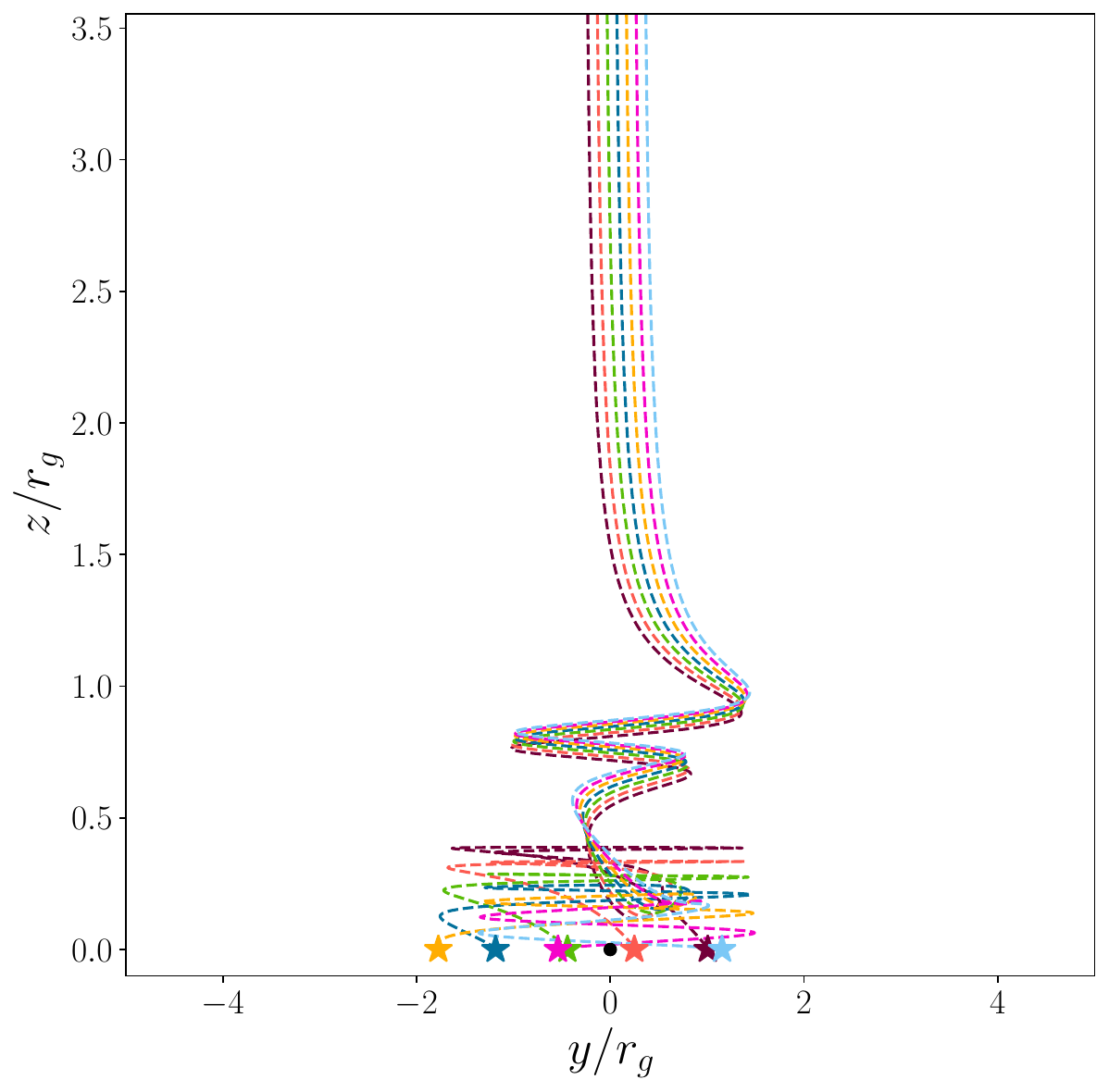}
\caption{Upper: Image plane views of Kerr metric accretion flows. Left: $a_\star = 1.01$, Right: $a_\star = 0.99$. These systems are observed at an angle $ \theta_{\rm inc} = 80^\circ$ from the singularities spin axis. While nearly identical in most of the image plane, a central cluster of photons is observed for $a_\star = 1.01$ which do not escape to the observer for $a_\star = 0.99$. Colorbar denotes the energy shifts of the photons over their trajectories, $f_\gamma \equiv E_{\rm obs}/E_{\rm emit}$.  Lower: The trajectories of the photons from the central cluster of the $a_\star = 1.01$ image plane (upper left panel)  projected into the $x-y$ (left) and $y-z$ (right) planes.  }
\label{RT4}
\end{figure*}

The physical reason that there are photon trajectories that propagate back out into the disc plane without being attracted to the singularity at $r = 0$ is related to the disappearance of the photon {prograde} circular orbit radius for super-extremal Kerr metrics. The radius of the Kerr metric's  circular photon orbit (denoted $r_p$) is given by the solution of the following cubic equation
\beq
x_p^{3/2} - 3x_p^{1/2} + 2a_\star = 0, \quad x_p \equiv r_p / r_g.
\eeq
For $a_\star > 1$, this cubic has no real solutions, while for $a_\star < -1$, the solution is \citep{Chandrasekhar83}
\beq
x_p = 4 \cosh^2\left({1\over 3} \cosh^{-1} \left[- a_\star \right] \right) .
\eeq
In the equatorial plane, only those photon orbits with pericentre radii smaller than $r_p$ terminate at the metric's singularity, or in other words, all equatorial photon orbits in the (prograde) super-extremal  Kerr metric escape to $r \to \infty$ at late times. Out of the equatorial plane, a small number of prograde photons hit $r=0$ (there is an analogous ``spherical'' photon orbit \citealt{Teo03}, which is non-zero but small for $a_\star > 1$) but the vast majority propagate back out into the disc plane (Fig. \ref{RT4}).

This property of super-extremal Kerr metrics results in the observer's image plane properties of Kerr discs being fundamentally discontinuous at $a_\star = 1$. This is because a cluster of photons  in the centre of the observers image plane, which would ordinarily be lost into the event horizon, in fact escape whenever $a_\star > 1$, and result in a central image spot, without analogy in the sub-extremal Kerr metric. In Fig. \ref{RT4} we display image plane  views of accretion flows around $a_\star = 0.99$ and $a_\star = 1.01$ Kerr singularities, both observed at $\theta= 80^\circ$. Only those photons emitted from $r>r_I$ are included.

While nearly identical in most of the image plane, a central cluster of photons is observed for $a_\star = 1.01$ which do not escape to the observer for $a_\star = 0.99$. These photons originate from radii close the discs inner edge, and so are intrinsically high-energy.  The lack of detections of the observed properties of these photons would act as evidence for the existence of event horizons in nature. A more detailed view of the photon trajectories in this central cluster is presented in the lower panel of Fig. \ref{RT4}.





\section{Super-extremal Kerr solutions as tests of conventional physics} 
\subsection{X-ray spectra of super-extremal solutions} 
The specific flux density $F_E$ of the disc radiation, as observed by a distant observer at rest (subscript ${\rm obs}$), is given by
\beq
F_{E}(E_{\rm obs}) = \int I_E(E_{\rm obs} ) \, \text{d}\Theta_{{{\rm obs} }} .
\eeq 
Here, $E_{\rm obs} $ is the observed photon energy and $I_E(E_{\rm obs} )$ the specific intensity,  both measured at the location of the distant observer.   The differential element of solid angle subtended on the observer's sky by the disk element is $\text{d}\Theta_{{{\rm obs} }}$. 
Since $I_E/ E^3$ is a relativistic invariant \citep[e.g.,][]{MTW}, we may write
\beq
F_{E}(E_{\rm obs} ) = \int f_\gamma^3 I_E(E_{\rm emit}) \, \text{d}\Theta_{{{\rm obs} }},
\eeq 
where the photon energy ratio factor $f_\gamma$ is the ratio of $E_{\rm obs} $ to the emitted local rest frame photon energy $E_{\rm emit}$:
\begin{equation}\label{redshift}
f_\gamma(r,\phi) \equiv 
{E_{\rm obs} \over E_{\rm emit}} = {1 \over U^t} \left(1 + {p_\phi \over p_t} \Omega \right)^{-1} . 
\eeq
In this expression $p_\phi$ and $-p_t$ are the angular momentum and energy of the emitted photon.
For an observer at a large distance $D$ from the source, the differential solid angle into which the radiation is emitted is
\beq
 \text{d}\Theta_{{{\rm obs} }} = \frac{\text{d}\alpha \, \text{d} \beta}{D^2} ,
\eeq 
where $\alpha$ and $\beta$ are the impact parameters at infinity \citep{Li05}.  See e.g.,  {Fig.\  (\ref{RT4})}.  

\begin{figure*}
\includegraphics[width=0.49\textwidth]{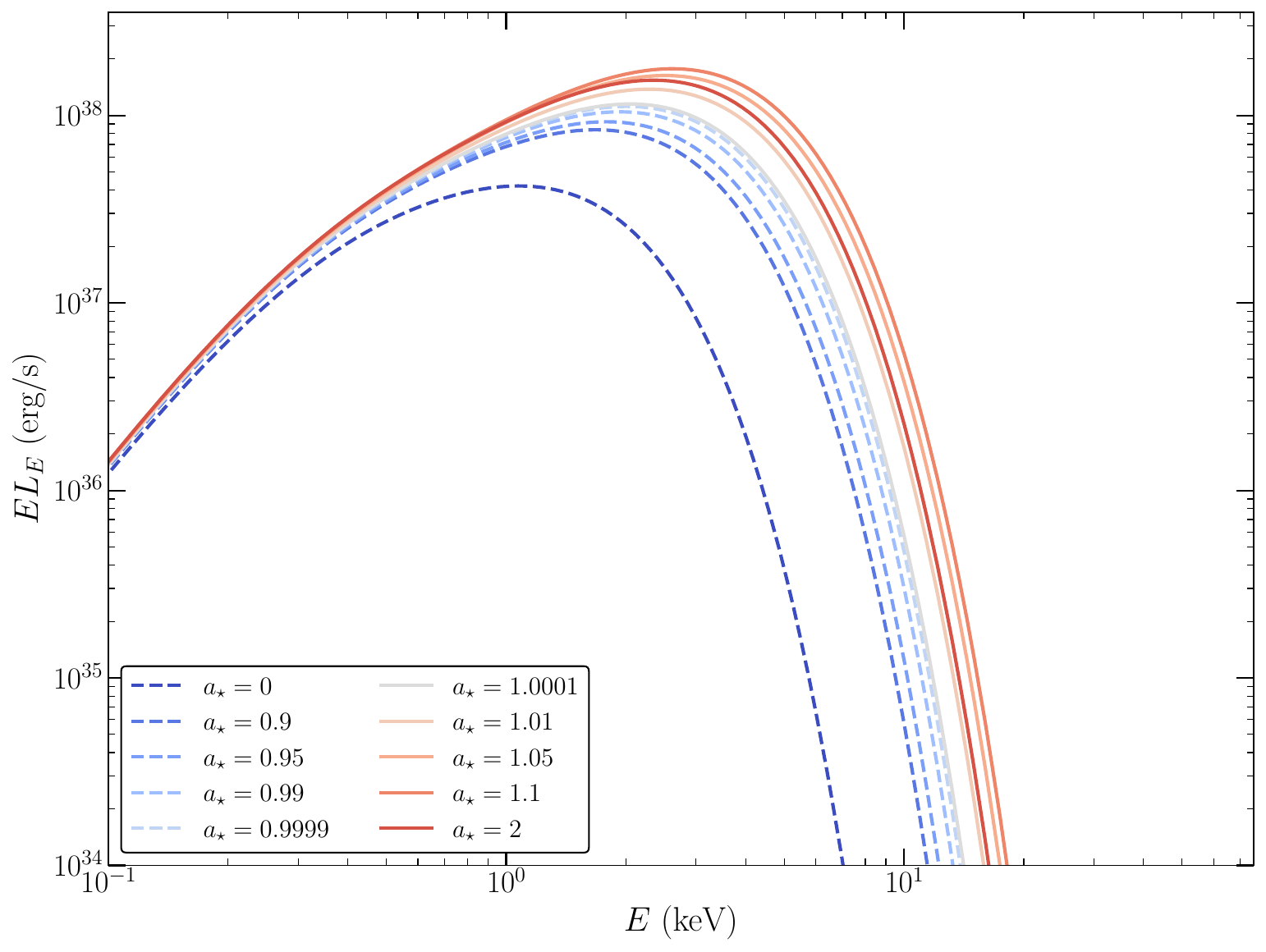}
\includegraphics[width=0.49\textwidth]{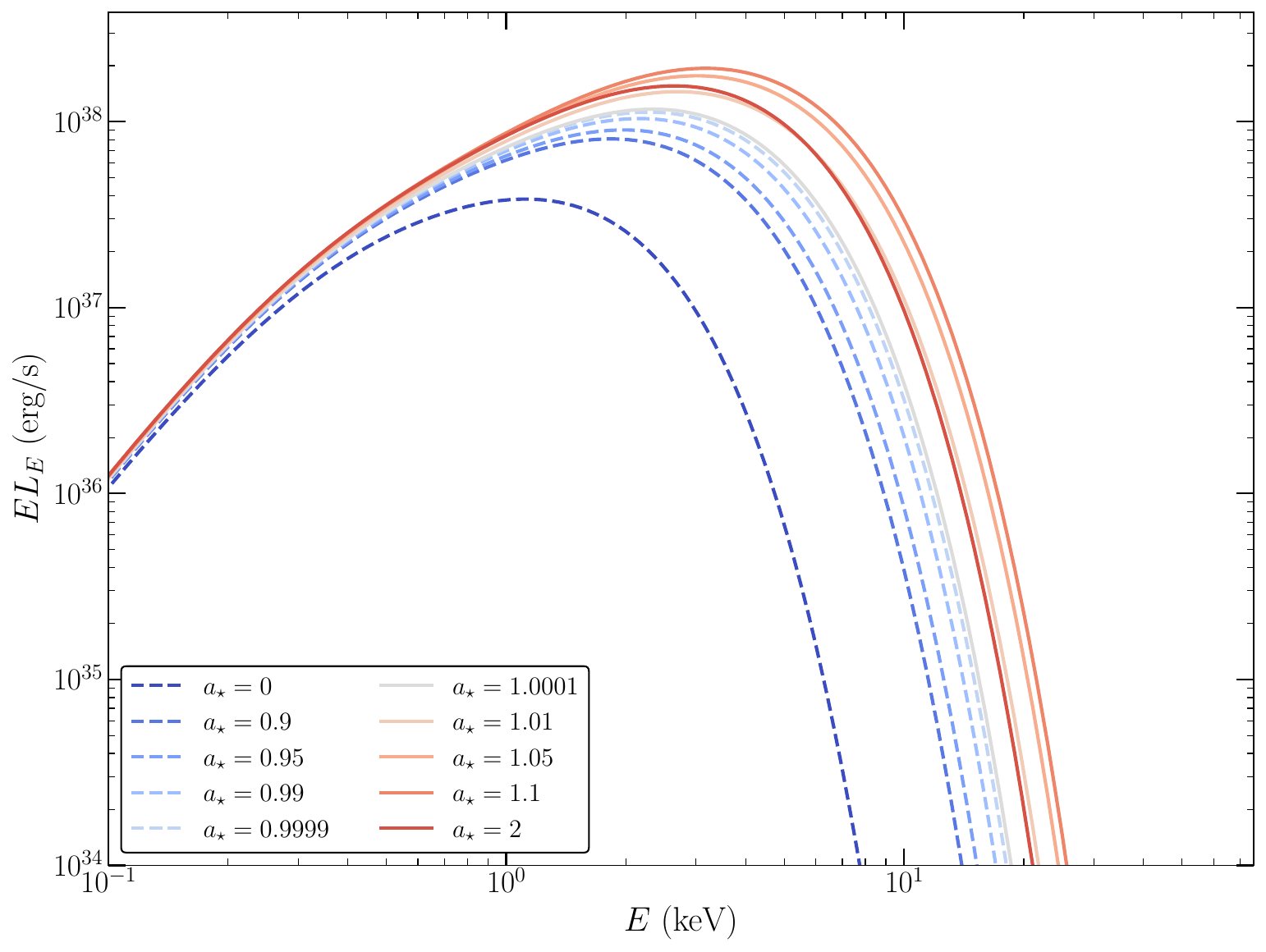}
\includegraphics[width=0.49\textwidth]{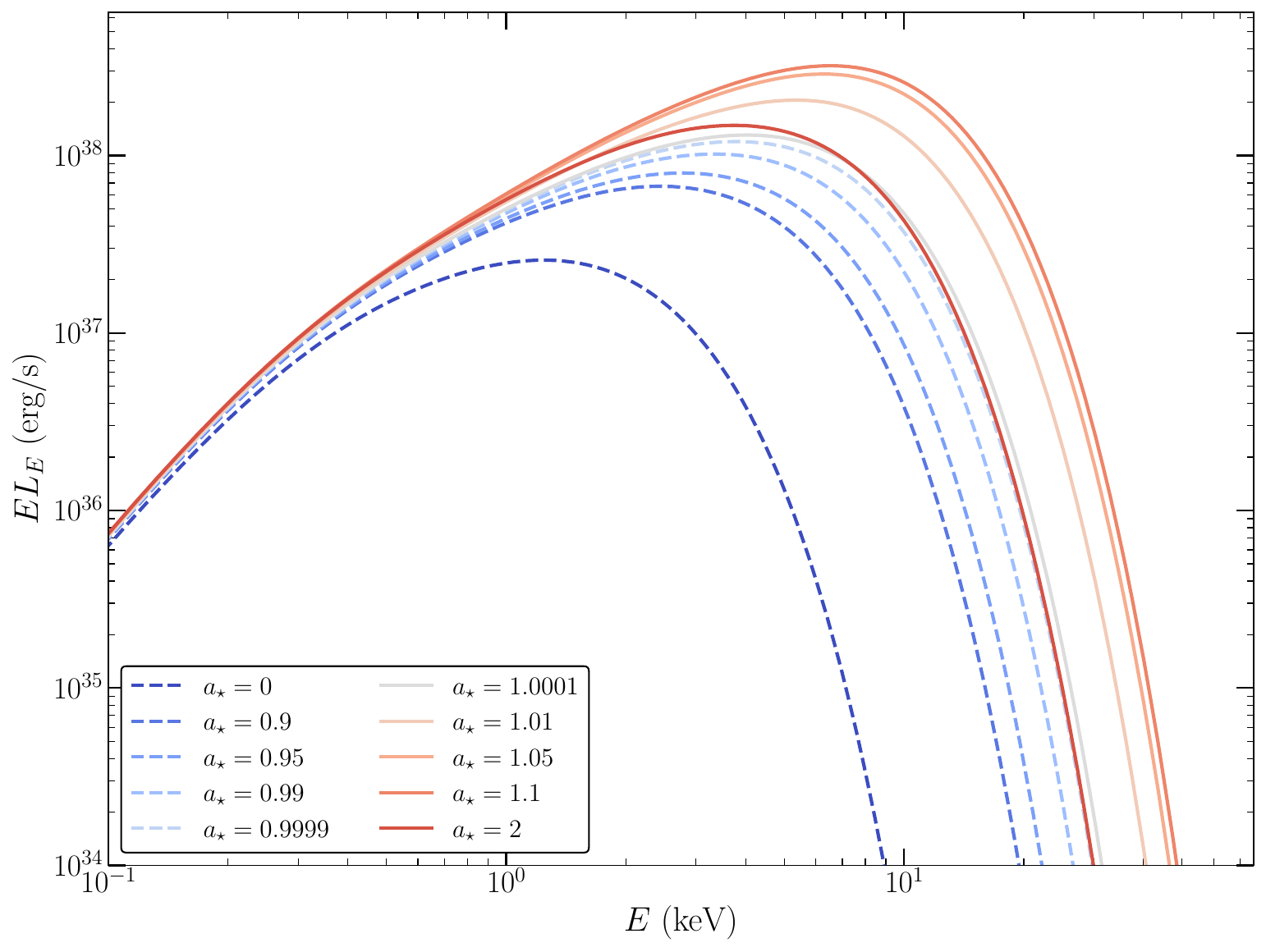}
\includegraphics[width=0.49\textwidth]{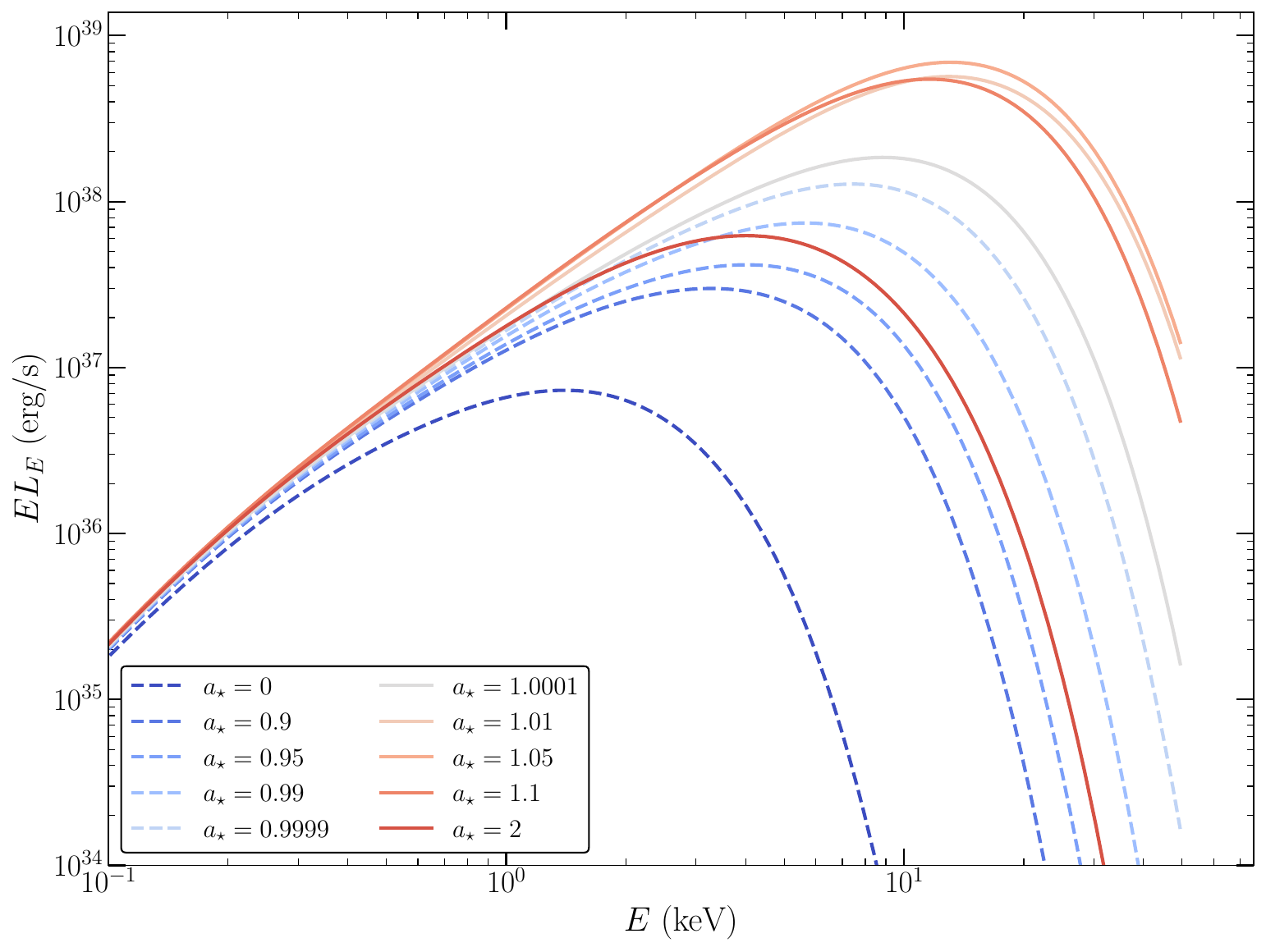}
\caption{X-ray spectra ($E L_E$) as a function of photon energy, from 0.1 to 50 keV, for Kerr metrics of different angular momentum parameters $a_\star$ (see Figure legend).  Dashed lines represent sub-extremal metrics, solid lines represent super-extremal metrics. The upper left panel is for a face on viewing angle  $ \theta_{\rm inc} = 5^\circ$, the upper right has $ \theta_{\rm inc} = 30^\circ$, the lower left has $ \theta_{\rm inc} = 60^\circ$ and the upper right $ \theta_{\rm inc} = 85^\circ$. The other system parameters are $M = 10 M_\odot$, $\dot M = 10^{15}$ kg/s.  We used the \citet{Done12} model for $f_{\rm col} (T)$. Each disc was modelled as having a vanishing ISCO stress. The outer edge of the disc was set at $ r_{\rm out} = 5000 r_g $.  }  
\label{spectra}
\end{figure*}

{In this paper we shall focus on the so-called ``soft state'' emission from an accreting source. Accreting X-ray sources are also observed in a wide range of other states, including the so-called ``hard'' state where emission is dominated by non-thermal emission, and the disc itself may not extend down to the ISCO. The model we develop here is not suitable for comparison to these harder states. } {Under this assumption, t}he specific intensity of the locally emitted radiation is given by a modified Planck function $B_E$, of the form 
\begin{align}\label{planck}
I_E(E_{\rm emit}) &= f_{\rm col}^{-4} B_E(E_{\rm emit},f_{\rm col} T) \\ &= \frac{2E_{\rm emit}^3}{h^2 c^2f_{\rm col}^4 } \left[ \exp\left( \frac{E_{\rm emit}}{k f_{\rm col} T} \right) - 1\right]^{-1} .
\end{align}
The colour correction factor, denoted $f_{\rm col}$, is included here so as to model the effects of radiative transfer in the disc atmosphere. 

In brief, while the surface temperature of the disc $T(r,t)$ is given by the constraints of energy conservation (e.g., equation \ref{temperature}), this temperature in fact corresponds physically to the temperature of the disc surface at a height above the midplane where the optical depth of the disc equals 1.  It is important to note, however, that the disc's central temperature is given  by \citep[e.g.,][]{Frank02} 
\beq
T_c^4 = {3\over 8} \kappa \Sigma T^4 ,
\eeq
where for standard astrophysical parameters $\kappa \Sigma \gg 1$ \citep[e.g.,][]{SS73}. 
This result highlights that the energy of the disc photons produced in the disc midplane is higher than the surface temperature, taking a value roughly $E_\gamma \sim kT_c$. Only if the liberated disc energy can be fully thermalised in the disc atmosphere do the photons emerge with temperature $T$ (eq. \ref{temperature}). On their path through the disc atmosphere, photons can either be absorbed and re-emitted (thus totally thermalising their energy), or they can undergo elastic scattering. Elastic scattering however, by definition, does not change the energy of the photon, and so if this process dominates in the disc atmosphere, photons will be observed to have the “hotter” temperatures associated with the altitudes closer to the disc midplane, not the disc’s $\tau = 1$ surface. This modifies the emergent disc spectrum, a result which is typically modelled with a so-called colour-correction factor $f_{\rm col}$, which  can be thought of as quantifying the relative dominance of these two different opacities in the disc atmosphere.

Note that the normalising factor $1/f_{\rm col}^{4}$ here ensures that, despite the temperature of  the emission being modified by the colour-correction factor, the total (integrated over all energies)  emitted luminosity remains $\sigma T^4$ (and therefore energy is conserved).  The value of the colour-correction factor $f_{\rm col}$ in general depends on the local properties of the emitting region \citep{Davis06}. In this work we use the \cite{Done12} model for $f_{\rm col}(T)$. 

The observed specific flux from the disc surface ${\cal S}$ is therefore formally given by
\beq\label{flux}
F_E(E_{\rm obs}) = {1\over D^2} \iint_{\cal S} {f_\gamma^3 f_{\rm col}^{-4} B_E(E_{\rm obs} /f_\gamma , f_{\rm col} T)}\,  {\text{d}\alpha\, \text{d} \beta} .
\eeq
Note that $f_\gamma$ will generally depend upon $\alpha$ and $\beta$.   With $T$ given by equation (\ref{temperature}),  ray tracing calculations determining  $f_\gamma(\alpha ,\beta)$, and a choice for the colour-correction made, the observed spectrum may be obtained with all relativistic effects (kinematic and gravitational Doppler shifts, and gravitational  lensing) included.  The isotropic X-ray luminosity $L_E\equiv 4\pi D^2 F_E$, is then given explicitly  by
\beq\label{lum}
L_E= 4\pi \iint_{\cal S} {f_\gamma^3 f_{\rm col}^{-4} B_E(E_{\rm obs} /f_\gamma , f_{\rm col} T)}\,  {\text{d}\alpha \,  \text{d} \beta} .
\eeq

In Fig. \ref{spectra}, we plot the observed X-ray spectra ($E L_E$ versus observing energy $E$) for sub and super-extremal Kerr discs observed at differing inclination angles. We focus on the Kerr spin ranges around the critical value $a_\star =1$.   Each spectrum is generated assuming $M = 10 M_\odot$ and $\dot M = 10^{15}$ kg/s. These are typical physical parameters for a Galactic X-ray binary.  We generate X-ray spectra for observed photon energies between 0.3 and 20 keV, which is typical of modern instruments. Clearly, the transition from sub to super-extremal Kerr metrics is a smooth and continuous from the perspective of the observed X-ray spectrum. 

\begin{figure}
    \centering
    \includegraphics[width=\linewidth]{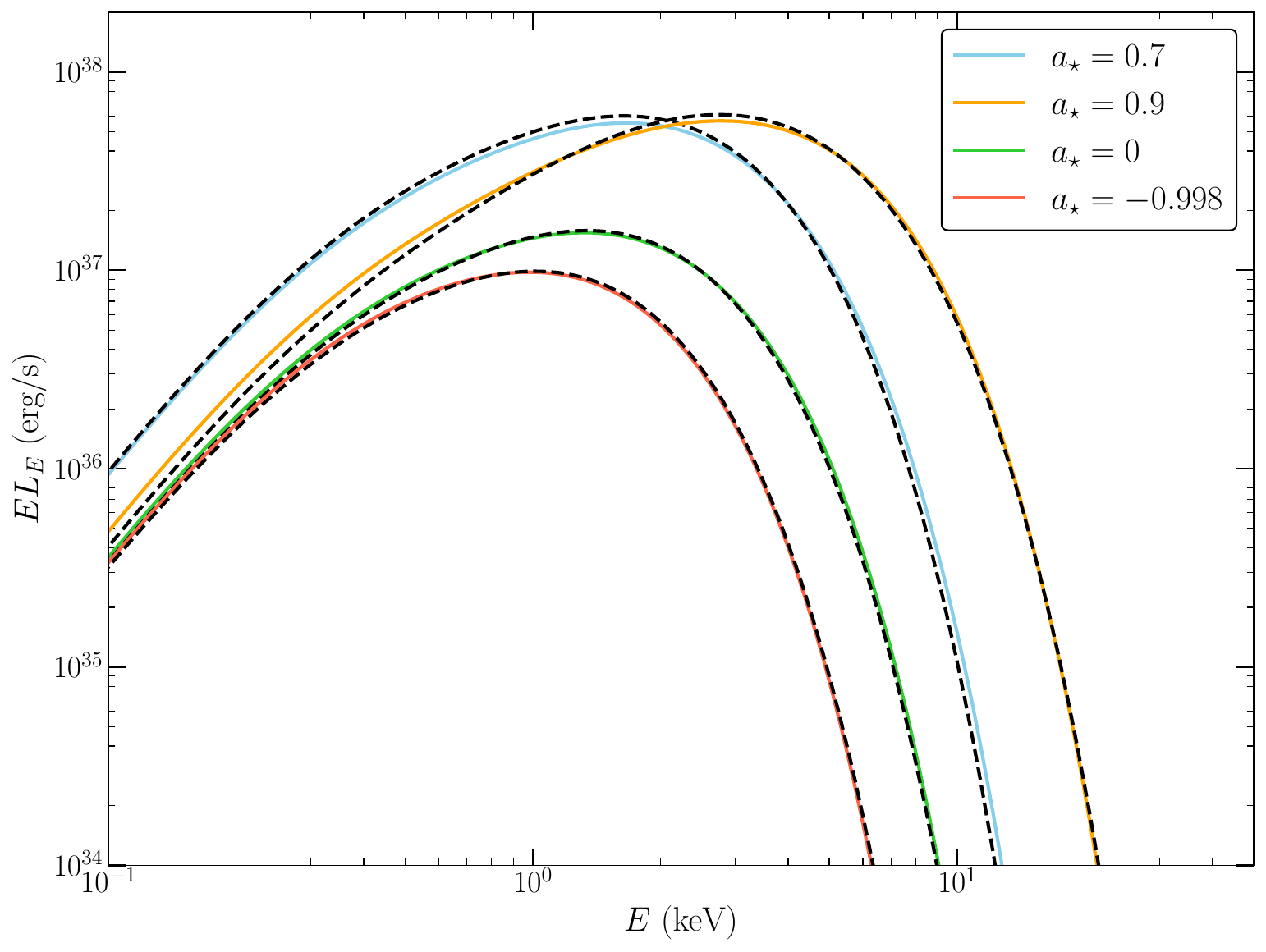}
    \caption{Examples of nearly-degenerate spectra between black hole discs $|a_\star| < 1$ and naked singularity disc solutions. Black hole discs are denoted by solid coloured curves, while naked singularity solutions are denoted by black dashed curves. The spins of the black hole discs are listed in the figure legend, and the nearly-degenerate naked singularity solutions have spin parameters (listed in the same order as the spins in the legend) of $a_\star = 4.23$, $a_\star = 3.2$, $a_\star = 8\sqrt{6}/3$ and $a_\star = 9$. These spin values have ISCO radii at very similar physical locations to their black hole counterparts. Only naked singularity spins in the range $1 < a_\star < 5/3$ are not degenerate with black hole discs.  }
    \label{fig:degenerate}
\end{figure}

{One property of naked singularity spacetimes which will be relevant for X-ray continuum fitting is the double valued nature of the ISCO radius as a function of the Kerr metric spin parameter $a_\star$ (e.g., Fig. \ref{fig1}). This, it transpires, results in nearly-degenerate X-ray spectra produced by black hole discs and naked singularity discs who share common ISCO locations, i.e., there is an observed degeneracy between black hole discs with spins $-1 < a_\star < 1$ and the naked singularity spin regime $5/3 < a_\star < 9$. This is highlighted in Fig. \ref{fig:degenerate}. In this figure we denote  black hole discs with solid coloured curves, while naked singularity solutions are denoted by black dashed curves. The spins of the black hole discs are listed in the figure legend, and the nearly-degenerate naked singularity solutions have spin parameters (listed in the same order as the spins in the legend) of $a_\star = 4.23$, $a_\star = 3.2$, $a_\star = 8\sqrt{6}/3$ and $a_\star = 9$. These spin values have ISCO radii at very similar physical locations to their black hole counterparts. In the following sections we shall demonstrate that only naked singularity spin parameters in the range $1 < a_\star < 5/3$ are not degenerate with black hole discs.   }

\subsection{Super-extremal discs as tests of conventional  disc theory}

A super-extremal Kerr metric describes a point mass with dimensionless angular momentum parameter $|a_\star| > 1$, and while remaining an exact solution of Einstein's field equations, exhibits a naked singularity and is therefore not expected to be present in nature. As such, if one allows super-extremal solutions to be part of the parameter space explored in spectral fitting, one would never expect to find a statistically acceptable fit which favours $a_\star > 1$ at high significance over all $|a_\star| < 1$ solutions. If however $a_\star > 1$ is favoured by the data then we must take seriously either (1) that conventional disc models are a poor description of astronomical sources, or (2) there are large and poorly understood systematic errors in our X-ray data collection and analysis, or (3) that either the cosmic censorship hypothesis or general relativity need revisiting. The null result of $a_\star < 1$ on the other hand would provide evidence that our conventional methods are successful. Including super-extremal Kerr discs in fitting procedures therefore represents a win-win addition to conventional approaches. 

For this proposal to represent a robust and meaningful test of theories of gravity and accretion, it is essential to firstly test for any intrinsic degeneracies
between black hole and naked singularity Kerr discs, before testing for the effects of adding additional physics. In other words, it is important that in the absence of additional physics the input parameters are reliably recovered from a vanishing ISCO stress Kerr disc.

\begin{table}
    \renewcommand{\arraystretch}{2}
    \centering 
    \begin{tabular}{|p{2.0cm} p{2cm} p{2cm}|}
    \hline
       Parameter  & Units & Allowed range  \\
    \hline 
       $M$ & $M_\odot$ &  $0 < M$ \\ 
       $a_\star$ & & $-\infty < a_\star < \infty$ \\
       $i$ & Degrees & $0 < i < 90$ \\ 
       $\dot M$ & $L_{\rm edd}/c^2$ & $0 < \dot M$ \\ 
       $\delta_{\cal J}$ & & $0<\delta_{\cal J}<1$ \\
       \hline 
    \end{tabular}
    \caption{The parameters of the \sk model,  their units and allowed ranges. The accretion rate parameter is defined in terms of the Eddington luminosity, which equals $L_{\rm edd} = 1.26 \times 10^{38} (M/M_\odot)$ erg/s.   }
    \label{skt}
\end{table}

To analyse this we generate ``fake'' data by using the {\tt fakeit} option in {\tt XSPEC} \citep{Arnaud96}. To be precise we simulate 10ks XMM observations of accretion flows in various different Kerr spacetimes, including the effects of the absorption of X-ray photons by neutral hydrogen between the disc system and observer with the {\tt tbabs} model \citep{Wilms00}. For the base disc spectrum we use the effective temperature derived above (eq. \ref{temperature}) for super-extremal metrics, and the classic \citet{PageThorne74, NovikovThorne73} disc solutions for sub-extremal metrics. We allow our disc solutions to have either a finite or vanishing ISCO stress. We use full ray-tracing calculations to include all relevant gravitational optics (lensing and energy shifting) effects.   We call this model {\tt superkerr}. The fake observations include the effects of the Poisson distribution of photon arrival times, the differing effective areas of the XMM telescope as a function of photon energy, and the finite probabilities that photons will have their energies mis-classified during the observing process. For an example ``fake'' 0.3-10 keV X-ray spectrum see Fig. \ref{example_spec}. 

With these fake X-ray spectrum generated we then fit this data with vanishing ISCO stress disc models, where the full spin parameter space is available.  To be precise, we fit the data by minimising the $\chi^2$ statistic of the model
where the parameters of the model are 
\beq
\Theta = (M, a_\star, \dot M, i, D_{\rm kpc}, N_{\rm H}) , 
\eeq
where $D_{\rm kpc}$ is the distance to the source in kiloparsec, and $N_H$ is the integrated  hydrogen column density of the galaxy along the line of sight from the source to the observer.  All parameters listed above are free to vary throughout the fitting procedure. See Table \ref{skt} for further details. 

We present some representative examples of the properties of sub- and super- extremal disc fits below.

\subsubsection{Case I: $M = 10M_\odot$, $a_\star = 0.7$, $i = 45^\circ$, $\dot M = 10^{15}$ ${\rm kg/s}$, $N_{\rm H} = 10^{22}$ ${\rm cm}^{-2}$, $D_{\rm kpc} = 10$ } 
For this example we demonstrate that for sub-extremal Kerr parameters, and a vanishing ISCO stress, standard fitting techniques recover the anticipated input parameters, but that a second naked singularity solution also provides an adequate description of the data. 

In the upper panel of Fig. \ref{example_spec} we show the fake data (denoted by black points and error bars) and the best fitting black hole solution (red dashed curve). The lower panel shows the ratio of the data and model.  The chi-squared of the best-fitting model $\chi^2_{\rm BH} = 1294.3$ is formally acceptable (for 1237 degrees of freedom), with best-fitting spin parameter $a_\star = 0.71$.

 However, in the lower plot (Fig. \ref{example_spec}) we again show the fake data (denoted by black points and error bars) and the a best fitting solution (red dashed curve), but this time for a naked singularity $a_\star > 1$ spacetime. The lower panel shows the ratio of the data and model.  The chi-squared of the best-fitting model $\chi^2_{\rm NS} = 1294.3$ is formally acceptable (for 1237 degrees of freedom), and is even marginally better than the black hole disc. This model has  best-fitting spin parameter $a_\star = 4.25$.  

Clearly there is a strong degeneracy between sub and super extremal Kerr X-ray spectra produced by discs with a vanishing ISCO stress. Physically this results from the double valued nature of the physical location of the ISCO radius as a function of Kerr spin parameter (Fig. \ref{fig1}). An X-ray spectrum is principally sensitive to the temperature profile of the hottest (innermost) disc regions, and in particular to the location of the ISCO. The two disc solutions shown in Fig. \ref{example_spec} have ISCO radii which are similar, explicitly  $r_I(0.7) = 3.34r_g$, and $r_I(4.25) = 3.43r_g$. 

We have verified that this degeneracy exists for all black hole solutions with a vanishing ISCO stress, namely there is always a naked singularity solution with spin parameter $5/3 < a_\star < 9$ with ISCO at the same physical location, and therefore a similar X-ray spectrum. 

This is an important result \citep[first discussed in][]{Takashi10}, and highlights two key points. The first is that simply ``measuring'' a spin parameter $5/3 < a_\star < 9$ does not necessarily imply a naked singularity solution, as it may simply be a degenerate solution for a black hole disc. A lack of any acceptable fits for $|a_\star| < 1$ is also required.  Secondly, the key spin parameter range to focus on is $1 < a_\star < 5/3$, as these solutions do not have parallel degenerate black hole solutions.   

\subsubsection{Case II: A rapidly rotating black hole and a finite ISCO stress }

\begin{figure}
\includegraphics[width=.49\textwidth]{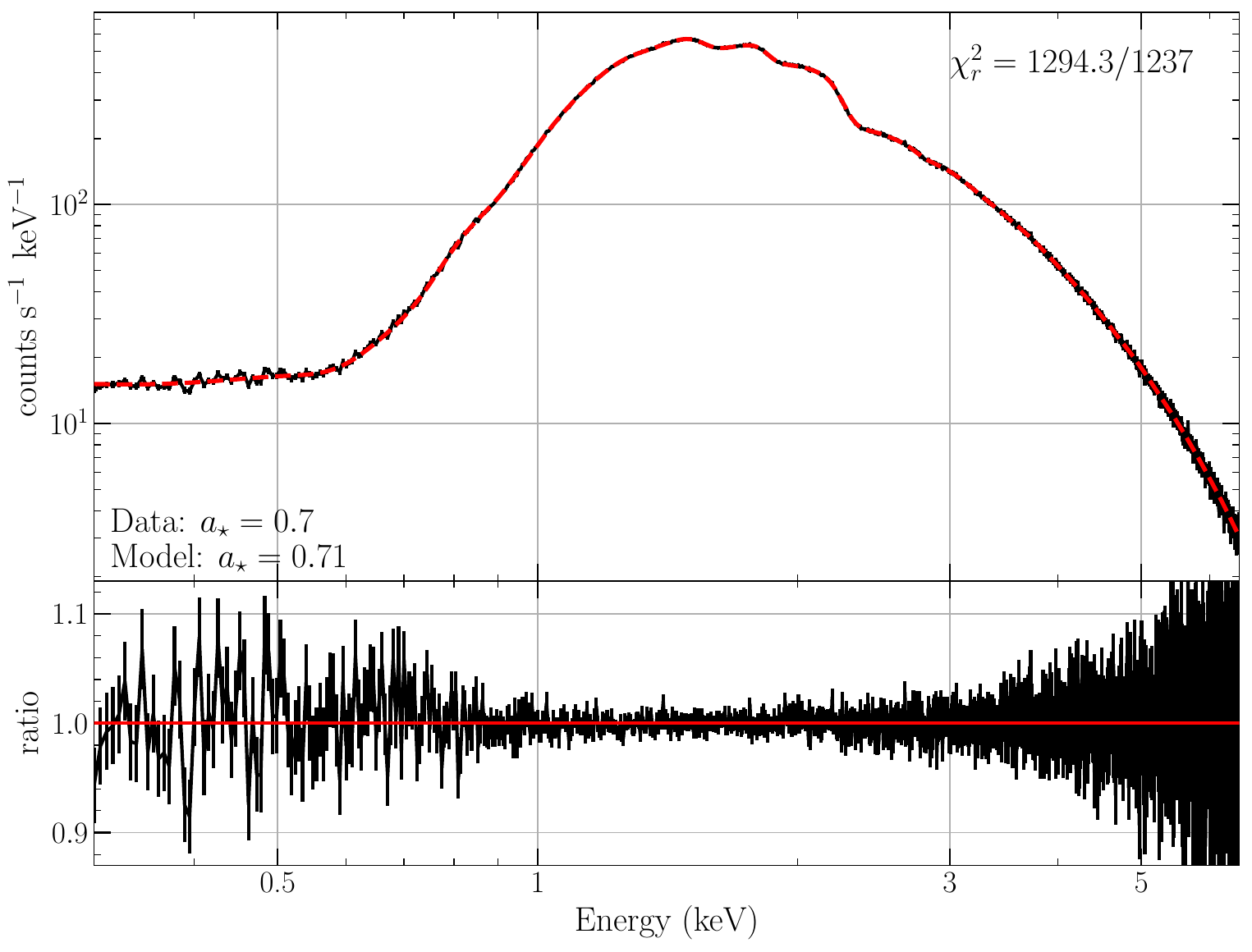}
\includegraphics[width=.49\textwidth]{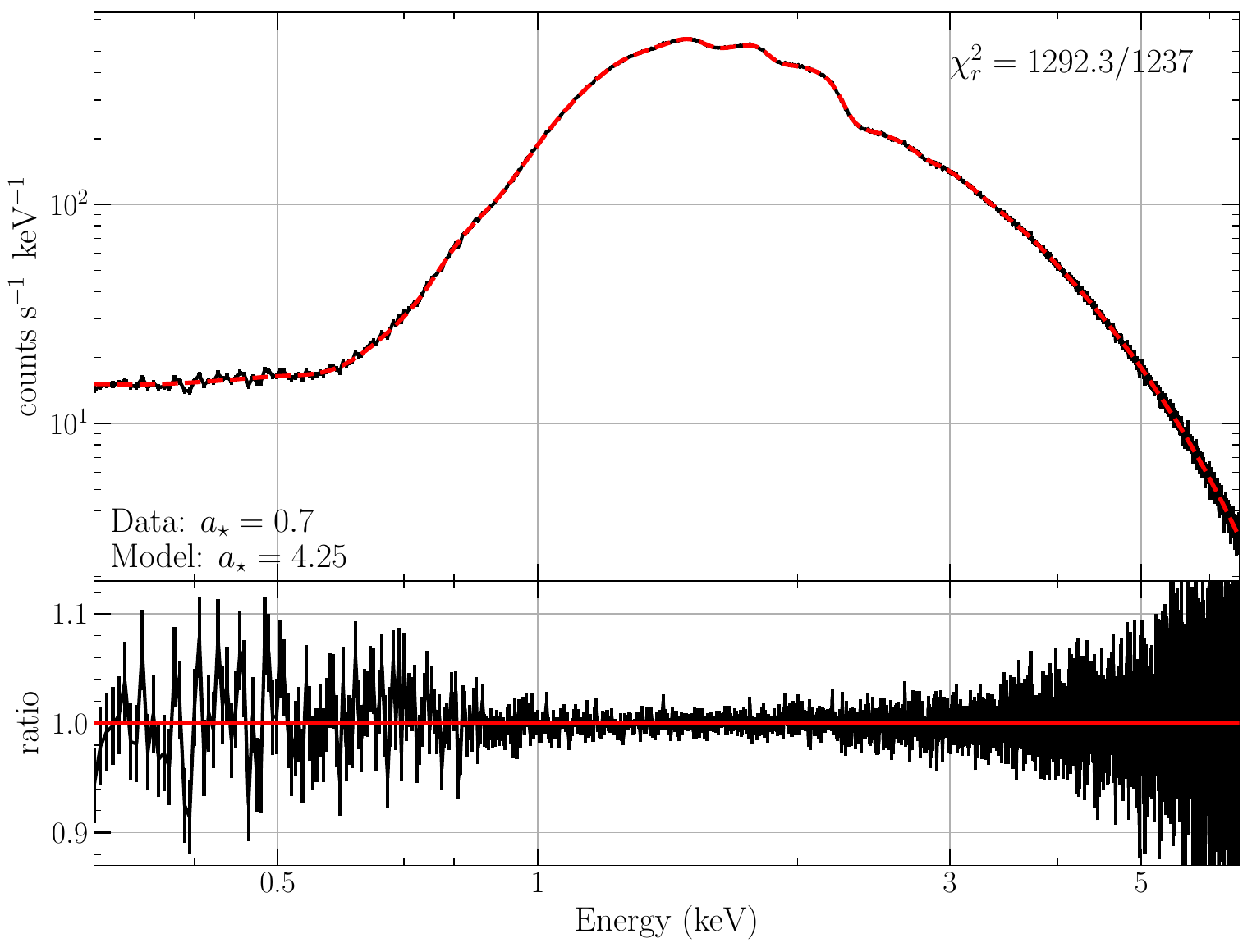}
\caption{An example ``fake''  X-ray spectrum, with inferred best fitting models. Top panels: the ``fake'' X-ray spectra (units of photons/s/keV; black points and error bars) and the best-fitting model (red-dashed curve). Bottom panels: the ratio between the data and the model. The fake data were produced with $a_\star = 0.7$, and the upper plot displays the best fitting black hole solution, with inferred spin of $a_\star = 0.71$. However, there is a naked singularity solution with $a_\star = 4.25$ (lower plot) which produces an equally good fit of the data ($\chi^2_{\rm BH} = 1294.3$, $\chi^2_{\rm NS} = 1292.3$). The ISCO radii of these two solutions are similar $r_I(0.7) = 3.34r_g$, $r_I(4.25) = 3.43r_g$. }
\label{example_spec}
\end{figure}

While spectra produced from a disc with a  vanishing ISCO stress are always well described by two vanishing ISCO stress solutions (one black hole and one naked singularity), when we add additional disc physics we break this degeneracy.   
We discuss now a key example of the effects of one such piece of additional physics, namely a non-zero ISCO stress.  

The notion that magnetic stresses could exert sizeable torques on the inner regions of accretion discs has a long history \citep{PageThorne74, Thorne1974, Gammie99, Krolik99}. Numerical studies over the past 20 years have generally shown that the magnetic fields that drive the MRI, the process thought to drive angular momentum transfer in discs, lead to extended evolutionary phases with non-zero stresses at the ISCO \citep[e.g.,][]{Shafee08, Noble10, Penna10, Zhu12, Schnittman16, Lancova19, Wielgus22}. The principal disagreement which now remains is the dynamical and observational importance of this non-zero ISCO stress, not its existence.

In a disc with a non-zero ISCO stress, angular momentum is transported {outwards} from the unstable disc region ($r < r_I$), back into the stable disc region ($r > r_I$). By the time they reach the event horizon, fluid elements (in numerical simulations) typically have angular momenta below what is required of a circular orbit at the ISCO (e.g., a 5--15 percent decrease was found in \citealt{Noble10}, while a 2 percent decrease was found in \citealt{Shafee08}). This liberated angular momentum slows the rotation of the inner edge of the stable disc region, and results in the presence of an angular momentum flux from the ISCO neighbourhood. Accretion flows in GRMHD simulations often have their hottest temperatures at the ISCO \citep[e.g.,][]{Zhu12, Schnittman16, Wielgus22}. Some observations of X-ray binaries hint at signatures of intra-ISCO emission, and are poorly described by conventional vanishing-ISCO stress models \citep{Fabian20}. 

Mathematically we allow for finite ISCO stress models in \sk by setting the angular momentum flux integration constant ${\cal F}_{\cal J}$ in equation (\ref{temperature}, see also eq. \ref{temp_def}) as 
\beq 
{\cal F}_{\cal J} = -(1 - \delta_{\cal J})\left[{\dot M r_g c \over 2\pi } \right] f(x_I), 
\eeq 
where $f(x)$ is the dimensionless integral 
\beq 
f(x) = \int^{\sqrt{r/r_g}}  {x^4 - 6x^2 - 3a_\star^2 + 8a_\star x \over x^4 - 3x^2 + 2a_\star x }\, {\rm d}x ,
\eeq 
with mathematical solutions discussed above. Physically $\delta_{\cal J}$ corresponds to the fraction of angular momentum the fluid passes back to the disc upon crossing the ISCO, where a  value of $\delta_{\cal J} = 0$ corresponds physically to no communication between the disc and plunging region and a vanishing ISCO stress. Numerical simulations \citep[e.g.][]{Shafee08, Noble10} place $\delta_{\cal J}$ somewhere in the range $\delta_{\cal J} \sim 0.02 - 0.2$. 

Modelling of X-ray spectra near universally assume a vanishing ISCO stress ($\delta_{\cal J}=0$) and if astrophysical discs in reality have non-zero ISCO stresses this may introduce biases into the spin measurements of continuum fitting studies. 

\begin{figure}
\includegraphics[width=0.49\textwidth]{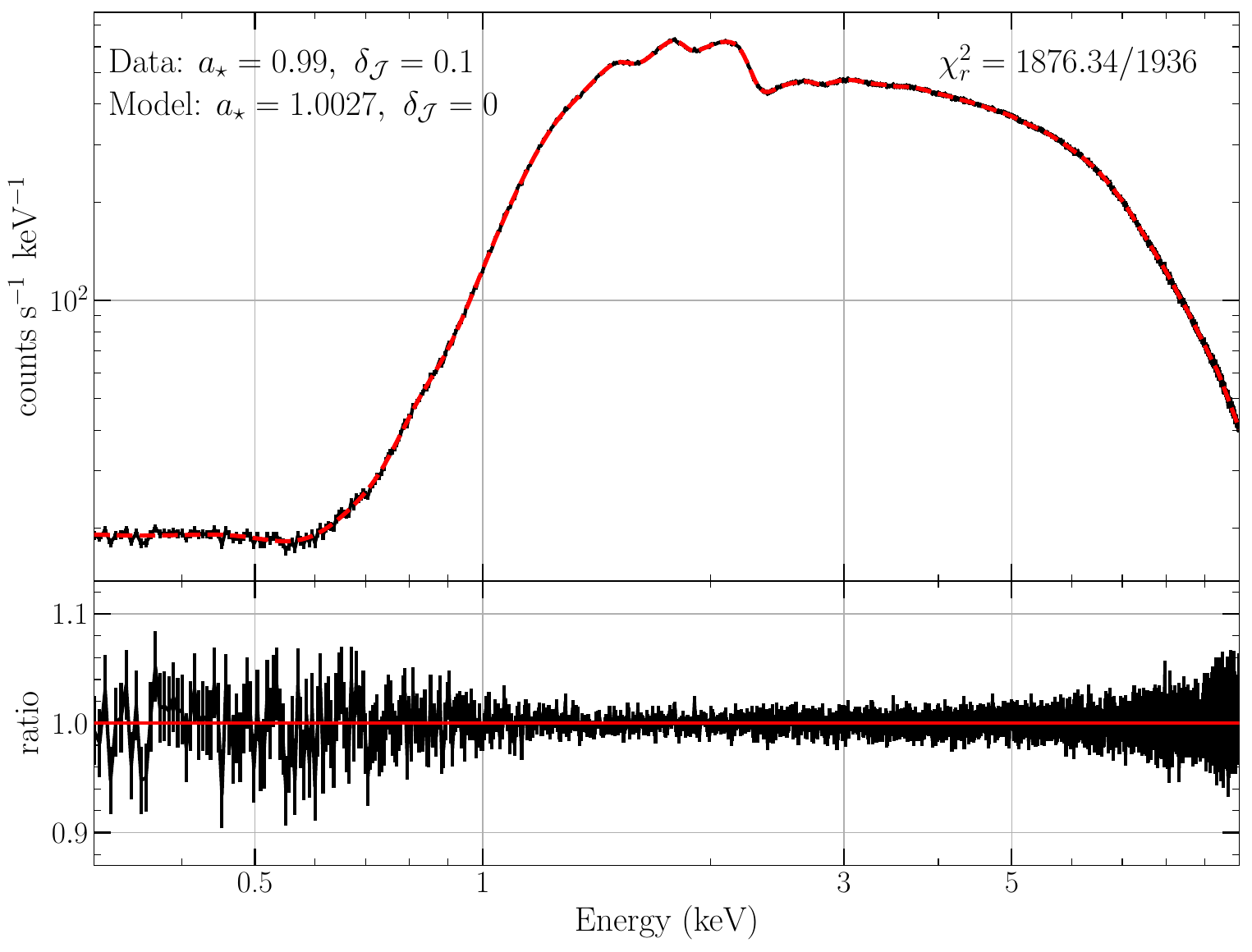}
\includegraphics[width=0.49\textwidth]{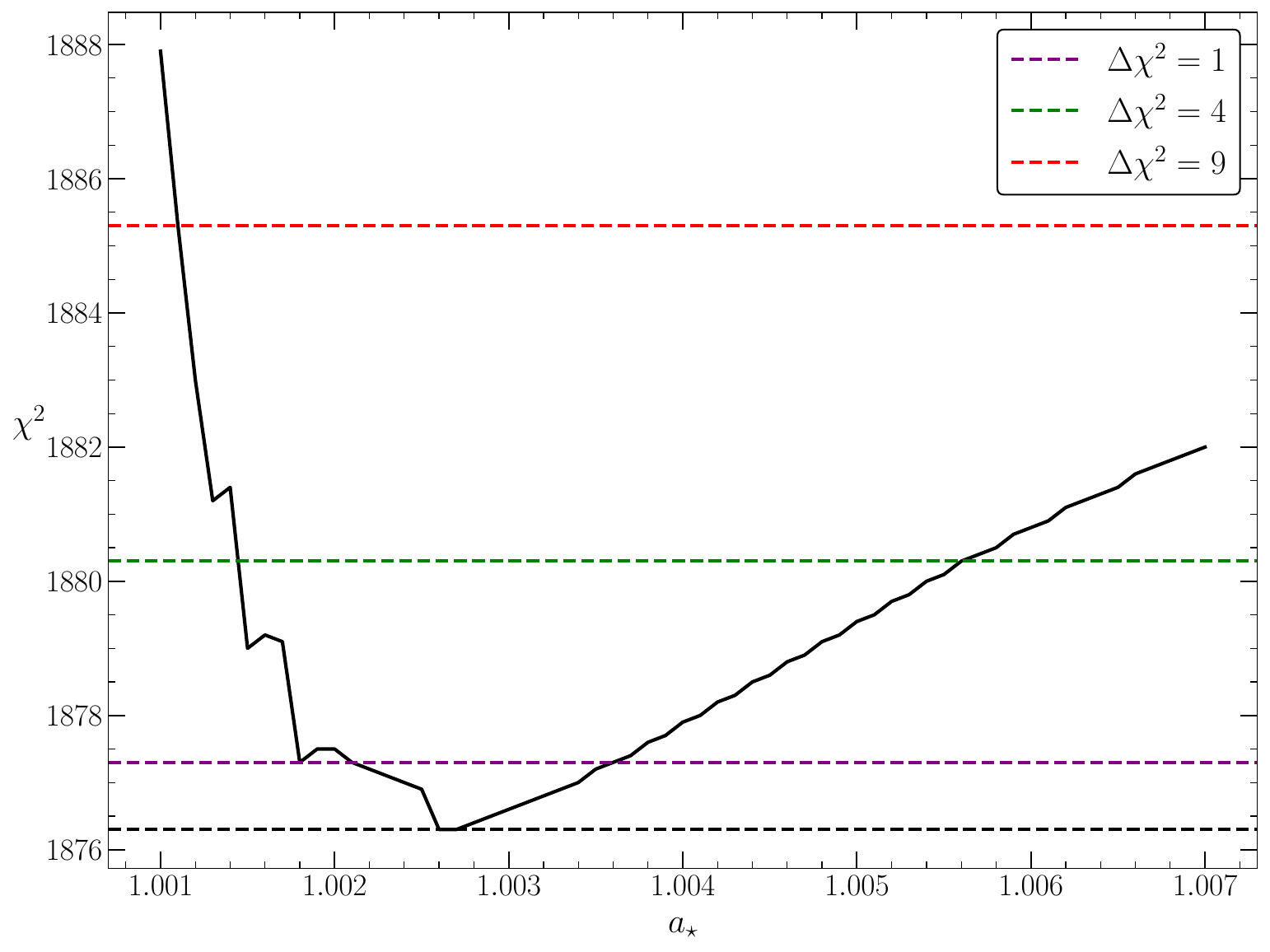}
\caption{An example of a super-extremal accretion flow solution which provides a formally excellent fit to the X-ray spectrum produced by a disc with a finite ISCO stress in the space time of a (sub-extremal) $a_\star = 0.99$ Kerr black hole. Top: the X-ray spectrum (units of photons/s/keV) of the fake data (black points and error bars) and the best-fitting model (red-dashed curve). Lower: the ratio between the data and the model. The best-fit is formally acceptable $\chi^2 = 1876.34$ for 1936 degrees of freedom. In the lower panel we show the $\chi^2$ fit statistic as a function of Kerr spin parameter, this best fit formally excludes black hole solutions ($a_\star < 1$) at high significance, and the spin is constrained to be $1.001 < a_\star < 1.01$ at 3$\sigma$.  }
\label{important_point}
\end{figure}

\begin{figure}
\includegraphics[width=0.49\textwidth]{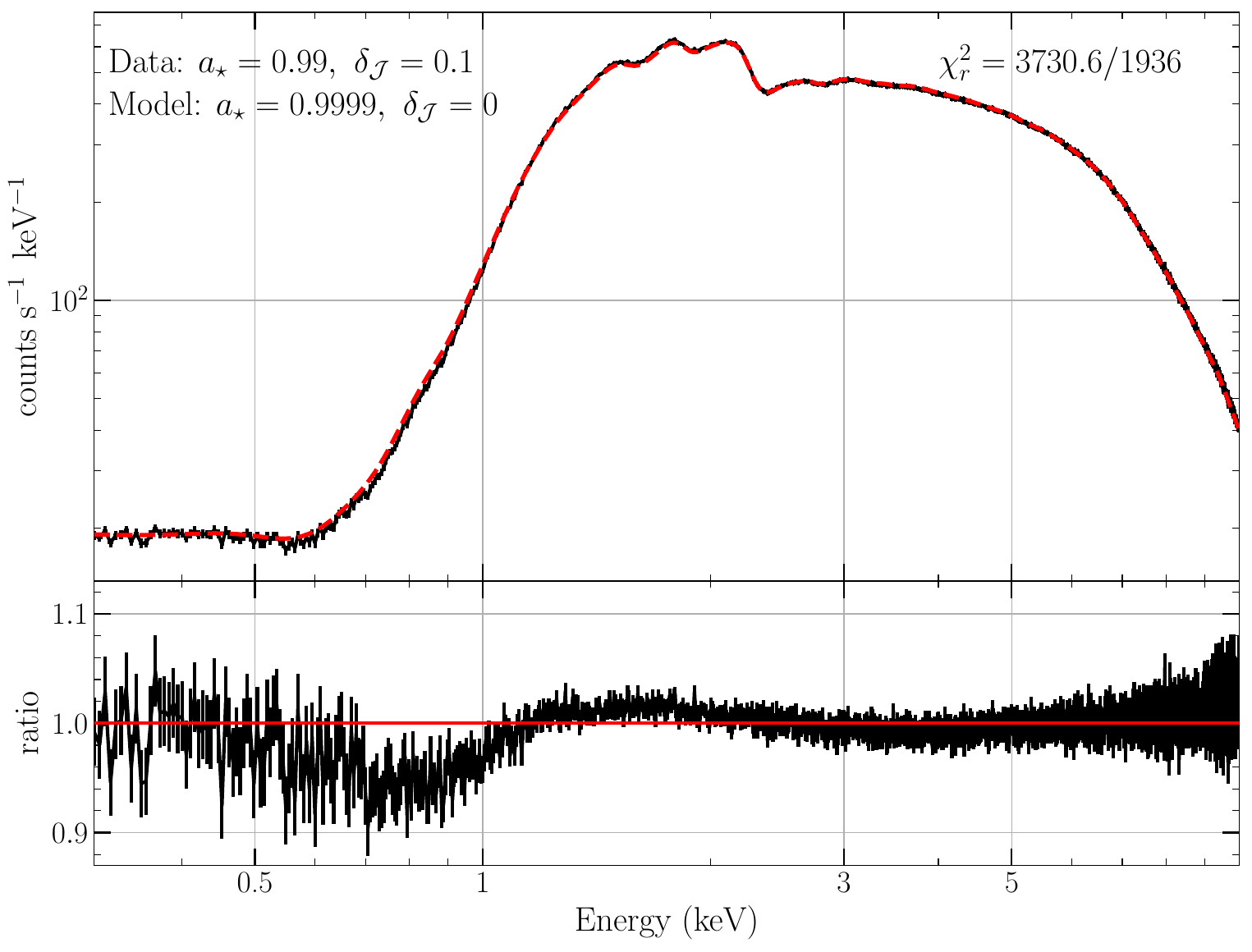}
\caption{The best-fitting black hole solution with a vanishing ISCO stress  (with best-fitting spin parameter $a_\star = 0.9999$) fit to the X-ray spectrum produced by a disc with a finite ISCO stress in the space time of a Kerr black hole with spin parameter $a_\star = 0.99$. Top panel: the X-ray spectrum (units of photons/s/keV) of the fake data (black points and error bars) and the best-fitting model (red-dashed curve). Lower: the ratio between the data and the model. The best-fit is not formally acceptable $\chi^2 = 3730.06$ for 1936 degrees of freedom, and is significantly poorer than a super-extremal disc (Fig. \ref{important_point}). This is due to vanishing ISCO stress disc X-ray spectra being insufficiently broad to describe a rapidly rotating black hole disc with a large finite ISCO stress. }
\label{important_point2}
\end{figure}

To demonstrate this we simulate a fake X-ray spectrum with the following intrinsic parameters: $M = 10M_\odot, a_\star = 0.99, \dot M = 10^{15}$ kg/s, $i = 80^\circ, D_{\rm kpc} = 10, N_{\rm H}= 10^{22} {\rm cm}^2$, and importantly a finite ISCO stress $\delta_{\cal J}=0.1$.   We then fit this fake data with an enforced vanishing ISCO stress model, and {all other} parameter{s} allowed to vary freely.   The resulting best fitting spin parameter was that of a super-extremal Kerr singularity, with spin $a_\star = 1.0027$ (Fig. \ref{important_point}). This best fitting model provided a formally excellent fit to the data ($\chi^2 = 1876.34$ for 1936 degrees of freedom). The requirement of a super-extremal spin is robust, with spins below $a_\star < 1$ strongly disfavoured, as can be seen by the $\chi^2$ statistic as a function of Kerr spin parameter $a_\star$ in Fig. \ref{important_point}.  

To further emphasise this point, we display in Fig. \ref{important_point2} the best fitting Kerr black hole disc model (we forced the spin parameter to remain below $a_\star \leq 0.9999$, as in the study of Cyg-X1 performed by \citealt{Zhao21}). The fit is poor ($\chi^2 = 3730.06$ for 1936 degrees of freedom) and is significantly poorer than for a super-extremal disc (Fig. \ref{important_point}). The best fitting spin parameter was pegged at our enforced limit $a_\star = 0.9999$. The poor fit was a result of black hole vanishing ISCO stress disc X-ray spectra being insufficiently broad to describe a rapidly rotating black hole disc with a large finite ISCO stress, as can be seen by the systematic residuals in the lower panel of Fig. \ref{important_point2}.

It is possible, although future studies will be required to verify this, that the very large spin values $(a_\star > 0.9985$ at 3$\sigma$, with a best fit of $a_\star = 0.9999)$ inferred by \cite{Zhao21} may be a result of neglecting a finite ISCO stress in the disc models used for spectral fitting.

\begin{figure}
\includegraphics[width=0.49\textwidth]{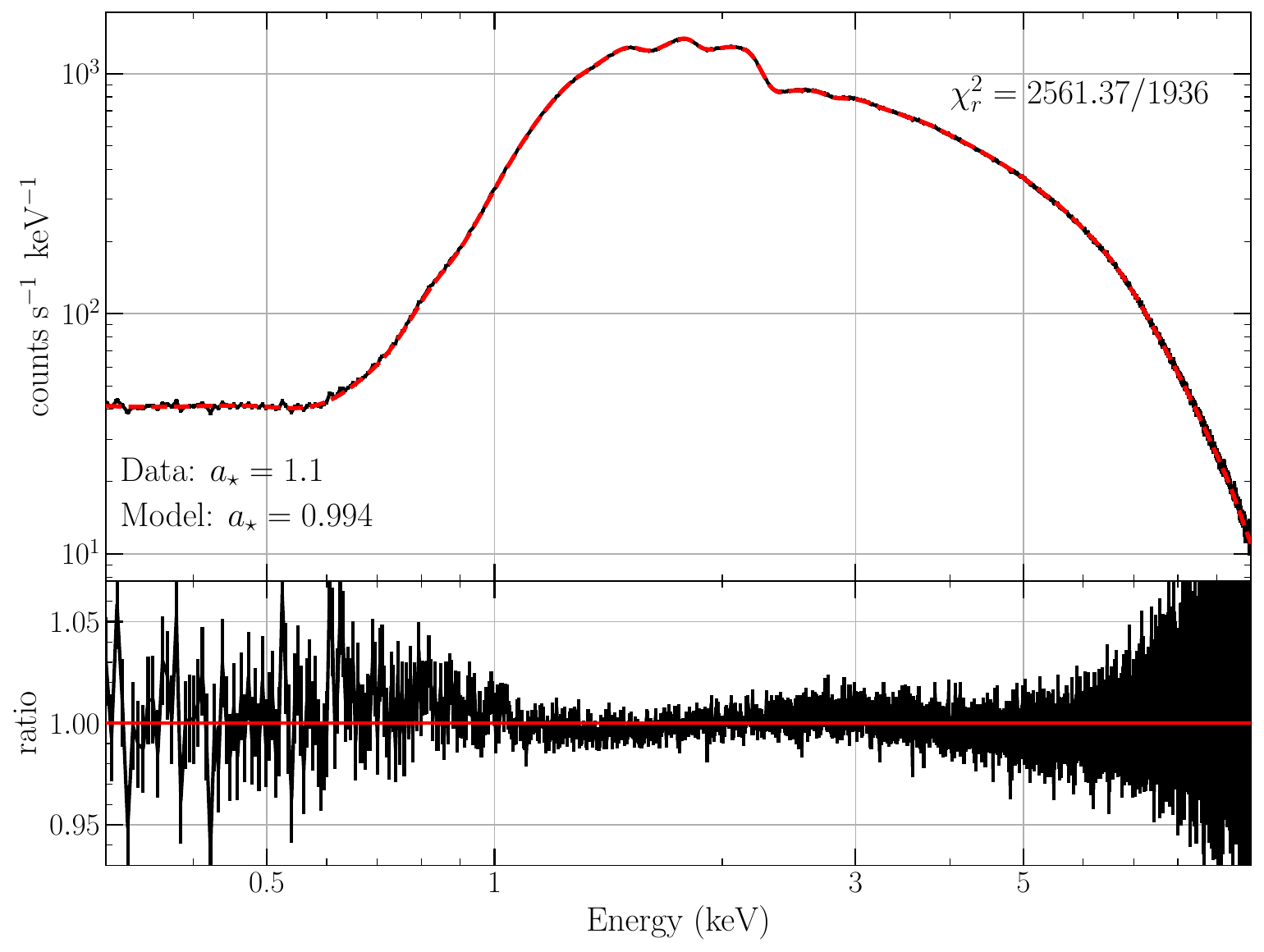}
\includegraphics[width=0.49\textwidth]{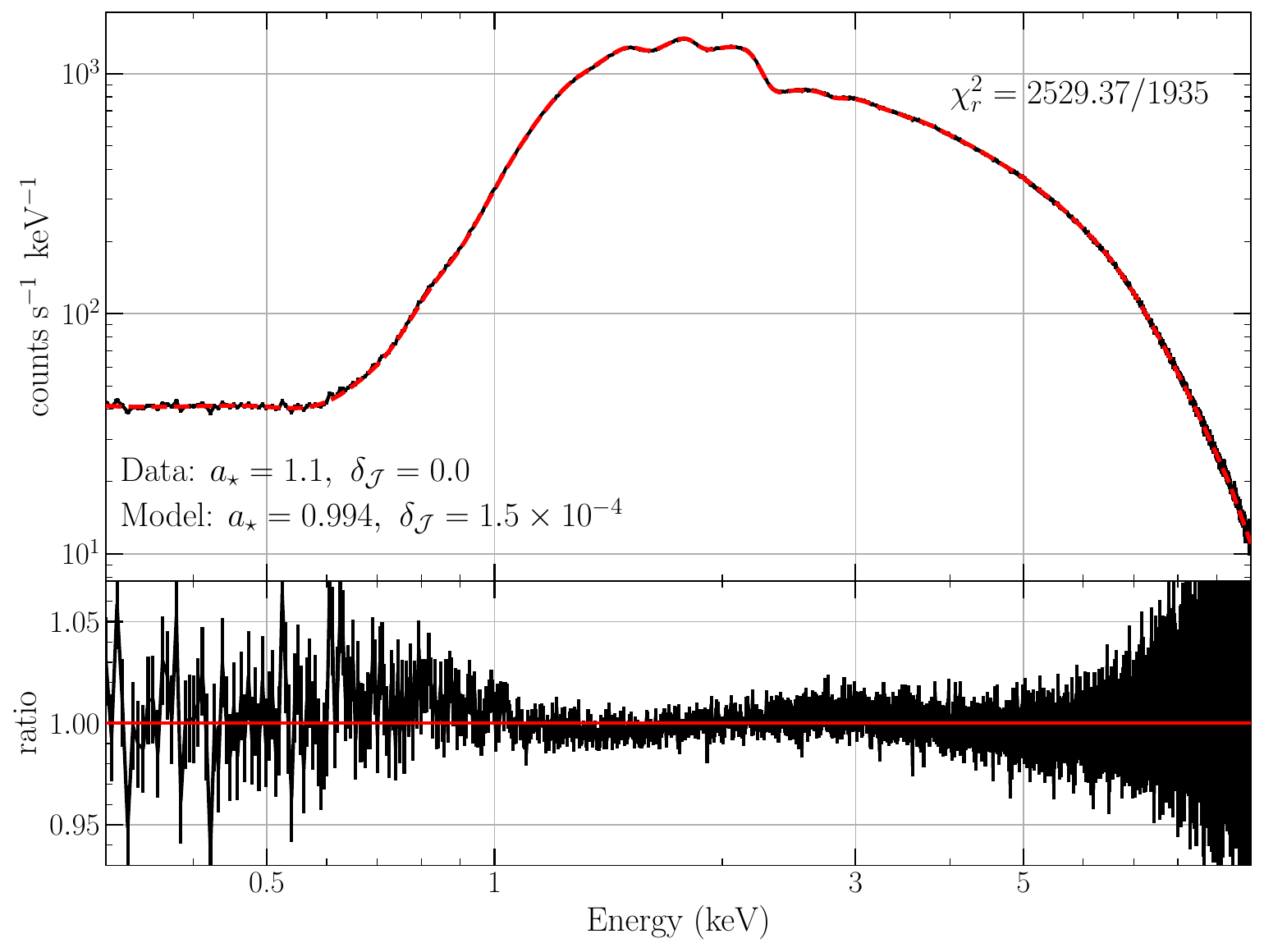}
\caption{Fits of vanishing ISCO stress (upper) and finite ISCO stress (lower) black hole disc models to a super-extremal vanishing ISCO stress disc model with $a_\star = 1.1$. Neither model provides an acceptable fit to the data, with the vanishing ISCO stress model resulting in $\chi^2 = 2561.37$, for 1936 degrees of freedom, and the finite ISCO stress model improving the fit slightly to $\chi^2 = 2529.13$, for 1935 degrees of freedom. Both black hole disc models inferred rapid rotation $a_\star > 0.99$.  }
\label{important_point3}
\end{figure}

\subsection{ Super-extremal discs as tests of theories of gravity }

We stress that while super-extremal discs provide formally acceptable fits of sub-extremal black hole discs with a finite ISCO stress, this degeneracy does not necessarily run in reverse.  To demonstrate this we produce a fake X-ray spectrum from a super-extremal Kerr disc, with spin parameter $a_\star = 1.1$. The other system parameters are: $M = 10M_\odot, \dot M = 10^{15}$ kg/s, $i = 45^\circ, D_{\rm kpc} = 10, N_{\rm H}= 10^{22} {\rm cm}^2$, and a vanishing ISCO stress $\delta_{\cal J}=0$. 

We then fit these spectra assuming a vanishing ISCO stress black hole disc, and do not find a statistically acceptable fit $\chi^2 = 2561.37$ for 1936 degrees of freedom (upper panel, Fig. \ref{important_point3}). Allowing the ISCO stress to also be a free parameter improves the fit somewhat, but not substantially $\chi^2 = 2529.13$, for 1935 degrees of freedom ($\Delta \chi^2 = 32.34$ for one extra parameter; lower panel, Fig. \ref{important_point3}).   This lack of a degeneracy between super-extremal and black hole discs holds broadly for spin parameters in the range  $1 \lesssim a_\star \lesssim 5/3$ (although for some spins close to these boundaries there is some degeneracy). 

This means that if there are indeed super-extremal Kerr naked singularities present in nature, X-ray spectral continuum fitting should be able to determine their presence provided their spin parameters are within the range $1 < a_\star < 5/3$. 

\section{Conclusions}
Motivated by recent spin measurements of Cyg-X1 obtained via the continuum fitting technique \citep{Zhao21}, we have produced an {\tt XSPEC} package \sk which produces valid disc emission profiles for {\it all} values of the Kerr metric spin parameter $a_\star$, and is not limited to just black holes $|a_\star | < 1$. The extension to Kerr metrics with $|a_\star| > 1$ involves the inclusion of metrics with naked singularities, which while remaining an exact solution of Einstein's field equations are not expected to be present in nature, a statement known as the weak cosmic censorship conjecture \citep{Penrose1969}. 

We remind the reader that there exists a natural degeneracy between black hole disc systems with $-1 < a_\star < 1$ and naked singularity disc systems in the restricted spin range $5/3 < a_\star < 9$, owing to their identical location of their ISCO radii in units gravitational radii (Fig. \ref{fig1}). However, over the interesting and relevant range $1 \lesssim a_\star \lesssim 5/3$ no such degeneracies exist for vanishing ISCO stress discs. 

This model will therefore allow the robustness of recent spin measurements obtained via the continuum fitting technique to be tested.  If spin parameters $1 < a_\star < 5/3$ are preferred by the data, we have demonstrated that this may be a signature of missing near-ISCO emission in standard black hole disc theory, and not necessarily the detection of a naked singularity (or equivalently a problem with our theories of gravity).  

On the other hand, most disc spectra produced by naked singularities (with spins $1 \lesssim a_\star \lesssim 5/3$) are completely without analogy in black hole discs, either with or without a finite ISCO stress. This new model therefore also allows for tests of our theories of gravity to be performed with X-ray data of astrophysical sources. 

To summarise the philosophy put forward in this paper, if one allows super-extremal solutions to be part of the parameter space explored in spectral fitting, one should never expect to find a statistically significant fit which favours $1< a_\star < 5/3$ at high significance over all $|a_\star| < 1$ solutions. If however $1 < a_\star < 5/3$ is favoured by the data then we must take seriously either (1) that conventional disc models are a poor description of astronomical sources (perhaps a missing ISCO stress, for example), or (2) there are large and poorly understood systematic errors in our X-ray data collection and analysis, or (3) that either the cosmic censorship hypothesis or general relativity need revisiting. The null result of $a_\star < 1$ on the other hand would provide evidence that our conventional methods are successful. Including super-extremal Kerr discs in fitting procedures therefore represents a win-win addition to conventional approaches.

\section*{Acknowledgments} 
 AM would like to thank  Pedro Ferreira and Georges Obied for stimulating discussions which initiated this work. This work was supported by a Leverhulme Trust International Professorship grant [number LIP-202-014]. For the purpose of Open Access, AM has applied a CC BY public copyright licence to any Author Accepted Manuscript version arising from this submission. AI acknowledges support from the Royal Society. This work is partially supported by the Hintze Family Charitable Trust and STFC grant ST/S000488/1.

\section*{Data availability}
No data was used in producing this manuscript. The {\tt XSPEC} model \sk is available at the following github repository:  \url{https://github.com/andymummeryastro/superkerr}

\bibliographystyle{mnras}
\bibliography{andy}

\appendix
\section{The solution of the temperature integral}\label{integral_app}
We wish to solve the integral
\beq
{\cal I}_a = {\dot M r_g c \over 2 \pi } \int^{\sqrt{r/r_g}}  {x^4 - 6x^2 - 3a_\star^2 + 8a_\star x \over x^4 - 3x^2 + 2a_\star x }\, {\rm d}x ,
\eeq
which can be approached  with standard partial fraction techniques. The approach, as was first shown in \citet{PageThorne74} for $|a| < r_g$, is to re-write the lower cubic in terms of its roots
\beq\label{roots}
x^3 - 3x + 2a_\star = (x - x_\alpha)(x-x_\beta)(x-x_\gamma) ,
\eeq
 before expanding the resultant fraction and integrating term by term. Start by expanding eq. \ref{roots} to determine three constraints on the roots of the cubic 
\begin{align}
&x_\alpha + x_\beta + x_\gamma = 0, \label{r1} \\
&x_\alpha x_\beta + x_\alpha x_\gamma + x_\beta x_\gamma = -3 , \label{r2}\\
&x_\alpha x_\beta x_\gamma = - 2a_\star . \label{r3} 
\end{align}
Equation \ref{r1} shows that we may write the roots of this cubic as 
\begin{align}
& x_\alpha = \xi , \\
& x_\beta = -{1\over 2} \xi + i \psi , \\ 
& x_\gamma = -{1\over 2} \xi - i \psi , 
\end{align}
where $\xi$ and $\psi$ are purely real numbers. Substituting these definitions into equation \ref{r2} relates $\xi$ and $\psi$ to one another 
\beq
\psi^2 = {3\over 4 }\xi^2 - 3 .
\eeq
Finally,  recall that $\xi$ itself, being a root of the cubic, must satisfy 
\beq
\xi^3 - 3\xi + 2a_\star = 0.
\eeq
The real root of the cubic is given by standard methods to be 
\beq
\xi = -2 \cosh\left[ {1\over 3} \cosh^{-1} \left({|a| \over r_g}\right)\right] ,
\eeq
and therefore 
\beq
\psi = \sqrt{3 \cosh^2\left[ {1\over 3} \cosh^{-1} \left({|a| \over r_g}\right)\right] - 3} . 
\eeq
With the properties of these roots determined, we can expand the in partial fractions and integrate.  The solution of the integral is formally
\beq\label{zeta_sol}
{\cal I}_a(x) = {\dot M r_g c \over 2 \pi } \left[ x - {3a_\star \over 2} \ln(x) + \sum_{\lambda = \{\alpha, \beta, \gamma\}} k_\lambda \ln\left(x - x_\lambda\right)\right] , 
\eeq
where 
\beq
 k_\lambda \equiv {2 x_\lambda - a_\star(1 + x_\lambda^2)  \over 2(1 - x_\lambda^2)} ,
 \eeq
and the sum over $\lambda$ should be interpreted as summing over each root. If one wishes to verify that  this is a purely real solution, one needs to consider the following combination of the two complex roots of the cubic ($x_\beta$, $x_\gamma$)
\beq\label{sum}
S = {2 x_\beta - a_\star(1 + x_\beta^2)  \over 2(1 - x_\beta^2)} \ln(x - x_\beta) + {2 x_\gamma - a_\star(1 + x_\gamma^2)  \over 2(1 - x_\gamma^2)} \ln(x - x_\gamma) .
\eeq
Note that 
\begin{multline}
\ln\left(x + {1\over 2}\xi \mp i \psi\right) = \ln\left(\sqrt{\left(x + {1\over 2 }\xi \right)^2 + \psi^2 }\right) \\ \mp i \arg \left({x + {1\over 2 }\xi} + i\psi \right) ,
\end{multline}
where $\arg(z)$ is the argument of a complex number, which is given explicitly in this case by 
\begin{multline}
\arg \left({x + {1\over 2 }\xi} + i\psi \right) =\tan^{-1} \left({\psi \over x + {1\over 2}\xi }\right), \\ = 2 \tan^{-1} \left({\psi \over x + {1\over 2}\xi + \sqrt{\left(x + {1\over 2 }\xi\right)^2 + \psi^2} }\right),
\end{multline}
where we have used the half angle formula 
\beq
\tan\left({1 \over 2 }\phi \right) = {\sin\phi \over 1 + \cos\phi} ,  
\eeq
to select the principal branch of the inverse tan function relevant for this solution. 
The combination 
\beq
k_\pm = {2 \left(-{1\over 2} \xi \pm i \psi\right) - a_\star\left[1 + \left(-{1\over 2} \xi \pm i \psi\right)^2\right]  \over 2\left[1 - \left(-{1\over 2} \xi \pm i \psi\right)^2\right]} ,
\eeq
can be split into real and imaginary parts 
\beq
k_\pm = {\cal R}\left[ k_\pm \right] + i {\cal I}\left[ k_\pm \right],
\eeq
where simple algebra reveals 
\begin{align}
{\cal R}\left[ k_\pm \right] &= {2 \xi \psi^2 (1 + {1\over 2} \xi a_\star) - \xi - a_\star \left(1 -\theta^2 \right) \over 2\left( \left( 1 -\theta\right)^2 + \xi^2 \psi^2\right)} , \nonumber \\
{\cal I}\left[k_\pm \right] &=  \pm {\xi \psi \left(\xi + a_\star  \left(1 + \theta  \right) \right) + 2 \psi (1 + {1\over 2} a_\star \xi) \left(1 - \theta  \right) \over 2\left( \left( 1 -\theta\right)^2 + \xi^2 \psi^2\right)} ,\nonumber  \\
\theta &\equiv  \left({1\over 4} \xi^2 - \psi^2 \right) = 3 - 2\cosh^2\left[ {1\over 3} \cosh^{-1} \left({|a_\star|}\right)\right].
\end{align}
The key point here is that both the imaginary parts of the logarithm, and the imaginary parts of the normalisation constants $k_\pm$, are proportional to $\pm
i$, and therefore the combination in equation \ref{sum} simplifies down to 
\begin{multline}
S = 2 {\cal R}\left[k_+\right] \ln\left(\sqrt{\left(x + {1\over 2 }\xi \right)^2 + \psi^2 }\right)  \\ + 4 {\cal I}\left[k_+\right] \tan^{-1} \left({\psi \over x + {1\over 2}\xi + \sqrt{\left(x + {1\over 2 }\xi\right)^2 + \psi^2} }\right),
\end{multline}
with all imaginary parts cancelling in the summation of the two terms. 

\subsection{Extremal $|a_\star| = 1$ limits} \label{extremal_temp_lims}
In the limit $|a_\star| = 1$ various quantities defined above vanish (i.e., $\xi = 0$ for $a_\star = \pm 1$), and some care is required in deriving the temperature profile.  In the extremal limit, the integral we wish to solve is
\beq
\left.{\cal I}\right|_{a_\star = \pm 1} = {\dot M r_g c \over 2 \pi } \int^{\sqrt{r/r_g}}  {x^4 - 6x^2 - 3 \pm 8 x \over x^4 - 3x^2 \pm 2 x }\, {\rm d}x .
\eeq
This integral has elementary solutions of the following form 
\beq
\left.{\cal I}\right|_{a_\star = \pm 1} = {\dot M r_g c \over 2 \pi } \left[ x \mp {3\over 2}\ln (x) \pm {3\over 2} \ln(x\pm 2) \right].
\eeq
The temperature profile can then be written explicitly as (assuming a vanishing ISCO stress)
\begin{multline}\label{e_t}
\left.\sigma T^4(r)\right|_{a_\star = \pm 1} = {3 G M \dot M \over 8 \pi r^3} \left({1 \over  1 - {3 / x^2} \pm {2 / x^3} }\right) \\ \Bigg[ 1 \mp {3 \over 2x} \ln\left({x\over x_I} \right) \pm {3\over 2x} \ln\left({x\pm 2\over x_I} \right) - {x_I \over x} \Bigg]  ,
\end{multline}
where $x_I(a_\star)$ is the square root of the ISCO radius in units of $r_g$
\beq
x_I(+1) = 1, \quad x_I(-1) = 3.
\eeq

\section{Ray tracing algorithm}\label{ART}
The isotropic luminosity of the disc emission is given by 
\beq\label{lum_app}
L_E= 4\pi  \iint_{\cal S} {f_\gamma^3 f_{\rm col}^{-4} B_E(E_{\rm obs} /f_\gamma , f_{\rm col} T)}\,  {\text{d}\alpha \,  \text{d} \beta} .
\eeq
where $f_{\rm col}$ and $T$ are, in principal, functions of disc radius.  The photon energy shift factor $f_\gamma$ is given by 
\beq
f_\gamma = \frac{1}{U^{0}} \left[ 1+ \frac{p_{\phi}}{p_t} \Omega \right]^{-1} .
\eeq
Here $U^t$ and $\Omega = U^\phi / U^t$ are 4-velocity components of the rotating  disc fluid, and $p_\phi$ and $p_t$ are photon 4-momentum components.  The ratio $p_\phi / p_t$ is a constant of motion for a photon propagating through the Kerr metric. As a conserved quantity, $p_\phi / p_t$  can  be calculated from the photon initial conditions.  

We assume a distant observer orientated at an inclination angle $i$ from the disc plane at a distance $D$. We set up an image plane perpendicular to the line of sight centred at $\phi = 0$, with image plane cartesian coordinates $(\alpha, \beta)$. A photon at an image plane coordinate  $(\alpha, \beta)$ has a corresponding spherical-polar coordinate $(r_i, \theta_i, \phi_i)$, given by  \citep{Psaltis12}
\begin{align}
r_i &= \left(D^2 + \alpha^2 + \beta^2\right)^{1/2}, \label{r0} \\
\cos\theta_i &=  r_i^{-1} \left( D \cos i + \beta\sin i\right) ,\\
\tan\phi_i &=   \alpha \left(D\sin i - \beta\cos i \right)^{-1}  .
\end{align}
The only photons which will contribute to the image have 3-momentum which is perpendicular to the image plane. This orthogonality condition uniquely specifies the initial photon 4-velocity \citep{Psaltis12}
\begin{align}
u^r_i &\equiv \left(\frac{\text{d} r}{\text{d}\tau}\right)_{\text{obs}} = \frac{D}{r_i} , \\
u^\theta_i &\equiv \left(\frac{\text{d} \theta}{\text{d}\tau}\right)_{\text{obs}}  = \frac{ D\left(D\cos i+ \beta\sin i \right) - r_i^2\cos i  }{r_i^2 \left( r_i^2 - \left(D \cos i   + \beta\sin i  \right)^2\right)^{1/2}} , \\
u^\phi_i &\equiv \left(\frac{\text{d} \phi}{\text{d}\tau}\right)_{\text{obs}} = \frac{- \alpha \sin i}{\left( D \sin  i - \beta\cos  i\right)^2 + \alpha^2} . \label{up0}
\end{align}
We note that the normalisation of these 4-velocity components can all be scaled by an arbitrary factor without effecting the trajectories (i.e., $\tau$ is an arbitrary affine parameter).

We trace the rays back from the observer to the disc by solving the null-geodesics of the Kerr metric (using the code \verb|YNOGK|, which is based on \verb|GEOKERR|: \cite{YangWang13, DexterAgol09}). 

Starting from a finely spaced grid of points $(\alpha, \beta)$ in the image plane, we trace the geodesics of each photon back to the disc plane, recording the location at which the photon intercepts the disc plane $(r_f)$, and the ratio $p_\phi/p_t$ for each photon.  The parameter $r_f$ allows the disc temperature $T$ to be calculated. The parameters $r_f$ and $p_\phi/p_t$ together uniquely define the energy-shift factor $f_\gamma$. The integrand  (of eq. \ref{lum_app}) can therefore be calculated at every grid point in the image plane, and the integral (\ref{lum_app}) is then calculated numerically. 


\label{lastpage}

\end{document}